\documentclass[12pt]{article}

\PassOptionsToPackage{
    colorlinks = true,
    linkcolor = HyperlinkBlue,
    citecolor = HyperlinkBlue,
    urlcolor = HyperlinkBlue
}{hyperref}

\usepackage[margin=1in]{geometry}

\usepackage[utf8]{inputenc}
\usepackage[T1]{fontenc}
\usepackage{newtxtext}
\usepackage{newtxmath}
\usepackage{microtype}

\usepackage{graphicx}
\usepackage{lscape}
\usepackage{xcolor}
\definecolor{HyperlinkBlue}{HTML}{0000FF}
\definecolor{OrcidGreen}{HTML}{6FA321}

\usepackage{hyperref}

\newcommand{\orcidlink}[2]{\href{#1}{\textcolor{OrcidGreen}{#2}}}

\usepackage{amsmath, amsfonts, afterpage}
\usepackage{bm}

\usepackage{amsthm}

\usepackage{tabularx}
\usepackage{longtable}
\usepackage{siunitx}
\usepackage{booktabs}
\usepackage{array}

\usepackage{caption}
\usepackage{subcaption}
\usepackage{appendix}
\usepackage{needspace}
\usepackage{soul}
\usepackage{setspace}

\usepackage[authoryear, round]{natbib}

\usepackage{autonum}

\providecommand{\bibcommenthead}{}

\newcommand{\supprefcolor}{HyperlinkBlue}
\newcommand{\supprefprefix}{Supplementary}
\newcommand{\suppautoref}[1]{\textcolor{\supprefcolor}{\supprefprefix}~\autoref{#1}}

\renewenvironment{abstract}{
  \vspace{1em}
  \noindent\textbf{Abstract}\vspace{0.3em}\\
  \noindent
}{
  \par\vspace{1em}
}

\newif\ifAnonymousManuscript

\AnonymousManuscriptfalse 

\newcommand{\BlindedOnly}[1]{%
    \ifAnonymousManuscript
    \else
        #1%
    \fi
}

\newcommand{\AnonymousOnly}[1]{%
    \ifAnonymousManuscript
        #1%
    \fi
}

\makeatletter
\renewcommand{\maketitle}{
    \begingroup
    \setlength{\parskip}{0pt}

    {\Large\bfseries \@title\par}
    \vspace{1.2em}

    \ifAnonymousManuscript

        {\normalsize Anonymous manuscript\par}

        \vspace{1.5em}

    \else

        {\normalsize
        Divan A. Burger$^{1,2}$,
        Sean van der Merwe$^{2}$,
        Emmanuel Lesaffre$^{3,4}$\par}

        \vspace{0.8em}

        \noindent
        \begin{tabularx}{\textwidth}{@{}p{1.2em}@{\hspace{0.4em}}X@{}}
        $^{1}$ & Cytel Inc., Waltham, MA, USA\\[2pt]
        $^{2}$ & Department of Mathematical Statistics and Actuarial Science, University of the Free State, Bloemfontein, Free State, South Africa\\[2pt]
        $^{3}$ & I-BioStat, KU Leuven, Leuven, Belgium\\[2pt]
        $^{4}$ & Department of Statistics and Actuarial Science, Stellenbosch University, Stellenbosch, South Africa
        \end{tabularx}

        \vspace{1.2em}

        \noindent
        \textbf{Correspondence:}
        Divan Burger, Cytel Inc., 1050 Winter Street, Waltham, MA 02451, USA.\\
        Email: \href{mailto:divanaburger@gmail.com}{\texttt{divanaburger@gmail.com}}

        \vspace{0.7em}

        \noindent
        \textbf{ORCID iDs:}
        Divan A. Burger, \orcidlink{https://orcid.org/0000-0001-8096-6371}{0000-0001-8096-6371};
        Sean van der Merwe, \orcidlink{https://orcid.org/0000-0002-8623-8639}{0000-0002-8623-8639};
        Emmanuel Lesaffre, \orcidlink{https://orcid.org/0000-0002-3747-6905}{0000-0002-3747-6905}.

        \vspace{0.6em}

    \fi
    \endgroup
}
\makeatother

\newcommand{\ManuscriptTitle}{Population-average trajectories and slopes in smooth random changepoint mixed models}
\title{\ManuscriptTitle}

\date{}

\begin{document}

    \maketitle

    \begin{abstract}
        Two-slope mixed-effects models with subject-specific changepoints are widely used to describe biphasic longitudinal trajectories. Treatment-arm summaries are often reported using a single ``typical subject'' curve obtained by plugging mean parameter values into the subject-specific model. This can be misleading when changepoint timing varies across individuals, because averaging subject-specific curves is not the same as evaluating one curve at average parameter values. A related issue is that individual trajectories may bend gradually rather than change slope through a literal kink, with transition sharpness varying between subjects. We propose population-average estimands for two-slope mixed models with random changepoints and smooth subject-level transitions. These estimands target the treatment-arm mean trajectory and instantaneous slope by averaging over fitted heterogeneity in changepoint timing and transition sharpness. We use a baseline-anchored normal-convolution soft hinge, which preserves the broken stick interpretation, includes the abrupt changepoint model as a limiting case, and allows analytic evaluation of the changepoint-marginalized kernels under a truncated normal changepoint distribution for fixed transition sharpness. Averaging over the fitted transition sharpness distribution is then performed by deterministic one-dimensional quadrature. In an application to HIV-1 RNA data from AIDS Clinical Trials Group Protocol~315, with censoring and heavy-tailed residuals, the marginal summaries reveal a more gradual population-level transition than conventional plug-in summaries. The marginal slope remains more negative for longer, and the marginal mean continues to decline after the plug-in curve has begun to flatten. These results illustrate the main reporting implication: in random changepoint models, representative subject curves should not be treated as population-average trajectories or slopes.
    \end{abstract}

    \vspace{0.6em}
    \noindent
    \begin{tabularx}{\textwidth}{@{}p{1.90cm}X@{}}
    \textbf{Keywords:} &
    Random changepoint $\boldsymbol{\cdot}$ Two-slope mixed model $\boldsymbol{\cdot}$ Population-average estimand $\boldsymbol{\cdot}$ Longitudinal biomarker $\boldsymbol{\cdot}$ Smooth hinge
    \end{tabularx}

    \newpage
    
    \section{Introduction}
    \label{sec:introduction}
    
    Two-slope mixed-effects models are commonly used for longitudinal biomarkers that show an early phase followed by a later phase with a different rate of change. Such biphasic patterns arise in settings including HIV-1 RNA under antiretroviral therapy \citep{perelson1996hiv}, eGFR under SGLT2 inhibition \citep{wanner2016empagliflozin}, tuberculosis bacterial burden \citep{donald2008early,burger2015bayesian}, and prostate-specific antigen after radiotherapy for localized prostate cancer \citep{bellera2008hierarchical}. In these models, each subject is typically described by an early slope, a later slope, and a changepoint around which the trajectory moves from one phase to the other \citep{laird1982random,davidian1995nonlinear,davidian2003nonlinear}. In biopharmaceutical studies, such two-phase longitudinal patterns often arise when early treatment response and later stabilization or decay are both of scientific interest, and treatment-arm summaries are commonly used to communicate biomarker trajectories over finite follow-up.
    
    When the changepoint is fixed in advance, or estimated as a common changepoint for all subjects, and the transition shape is fixed, summarizing the treatment-arm trajectory is relatively straightforward. In that setting, averaging over zero-mean subject-specific intercepts and slopes gives the same mean trajectory obtained by setting those random effects to zero and using the corresponding population-level intercept and slopes. The difficulty arises when the transition itself varies between subjects. In a random changepoint model, subjects may move from the early phase to the late phase at different times, so a representative subject curve that fixes the changepoint at a typical value is not generally equal to the average of the subject-specific curves. The same issue arises for instantaneous slope summaries: even if each subject changes slope abruptly, variation in changepoint timing can produce a smooth population-average slope. That population-level smoothness is different from subject-level smoothness, where an individual trajectory may itself bend gradually rather than change slope through a literal kink. The width of that subject-level bend may also differ between subjects. Thus, transition timing and transition shape are two sources of heterogeneity that can affect how treatment-arm trajectories and slopes should be summarized.

    In this paper, we define population-average summaries for two-slope mixed models with random changepoints and smooth subject-level transitions. The focus is not only on fitting subject-specific smooth transitions, but on deriving treatment-arm mean and slope summaries after averaging over heterogeneity in both changepoint timing and transition sharpness. The summaries of interest are the marginal mean trajectory and the marginal instantaneous slope, which describe the expected underlying trajectory and expected instantaneous rate of change for a randomly selected subject from the treatment arm, rather than the trajectory of a single representative subject. The subject-level transition is modeled using a baseline-anchored normal-convolution soft hinge, which preserves the broken stick interpretation, allows subject-specific transition sharpness, and recovers the abrupt changepoint model as a limiting case. A useful feature of this parameterization is that, conditional on the transition sharpness parameter, the changepoint-marginalized kernels can be evaluated analytically using normal and bivariate normal distribution functions. Thus, when transition sharpness is also random, evaluation of the marginal summaries reduces to a one-dimensional numerical average over transition sharpness rather than a two-dimensional numerical integration over both transition timing and transition sharpness. To the best of our knowledge, this baseline-anchored normal-convolution random changepoint formulation, together with the associated population-average mean and slope summaries, has not previously been developed for models with both random changepoint timing and random transition sharpness. The formulation is chosen specifically to preserve the broken stick interpretation while making the required marginalization tractable.

    This work is related to several strands of the literature on changepoint modeling. The Bacon-Watts formulation provides an early smooth transition between two linear phases \citep{bacon1971estimating}. Smooth random changepoint models have also been proposed for longitudinal trajectories \citep{hout2011smooth}, while sigmoidal mixed models provide another flexible way to describe gradual transitions over time \citep{capuano2018sigmoidal,capuano2023nlive}. Random changepoint mixed models have been developed for abrupt segmented trajectories, including both maximum likelihood and Bayesian formulations \citep{muggeo2014segmented,brilleman2017bayesian}. More recently, segmented regression with smooth transition has been considered in a random changepoint framework \citep{singer2021random}. These contributions mainly address how changepoints and smooth transitions can be represented and estimated at the subject level. Our focus is different: we define treatment-arm marginal mean and slope summaries for a model with both random changepoint timing and random transition sharpness, and use the normal-convolution parameterization to make the required averaging tractable, leaving only a one-dimensional numerical average over transition sharpness.
    
    We use longitudinal HIV-1 RNA data from ACTG~315 as an illustrative clinical-trial biomarker application with left censoring, heavy-tailed residual variation, and clear two-phase longitudinal behavior. The application illustrates that the proposed summaries can reveal a more gradual population-level transition than conventional plug-in curves, while the smooth changepoint formulation improves observation-level predictive performance relative to the abrupt changepoint model.
    
    The remainder of the paper is structured as follows. \autoref{sec:motivating-data} describes the motivating longitudinal HIV-1 RNA data. \autoref{sec:smooth-model} introduces the smooth random changepoint model. \autoref{sec:marginal-trajectories} defines the marginal estimands and derives tractable evaluation formulas. \autoref{sec:plugin-numerical} presents a numerical experiment comparing plug-in and marginal summaries using values calibrated to the motivating application. \autoref{sec:application} applies the method to the motivating antiretroviral therapy trial data. \autoref{sec:discussion} concludes with interpretation and extensions.

    \section{Illustrative application: HIV-1 RNA}
    \label{sec:motivating-data}
    
    We use longitudinal HIV-1 RNA measurements from AIDS Clinical Trials Group (ACTG) Protocol 315 as an illustrative application \citep{lederman1998immunologic,MORRIS2020}. ACTG~315 was an early combination antiretroviral therapy study with a single multidrug regimen. The dataset comprises 46 participants and 361 plasma HIV-1 RNA measurements (copies/mL). Participants contributed a median of 8 measurements each, including baseline (range: 4 to 10), over a maximum follow-up of approximately 6.5 months. Viral load is analyzed on the $\log_{10}$ scale. Measurements reported as below the lower limit of quantification (LLOQ) are treated as left-censored at 100~copies/mL, that is, at $\log_{10}\left(100\right)$; overall, 11.1\% of observations are censored.
    
\begin{figure}[!htbp]
      \centering
      \includegraphics[width=0.65\textwidth]{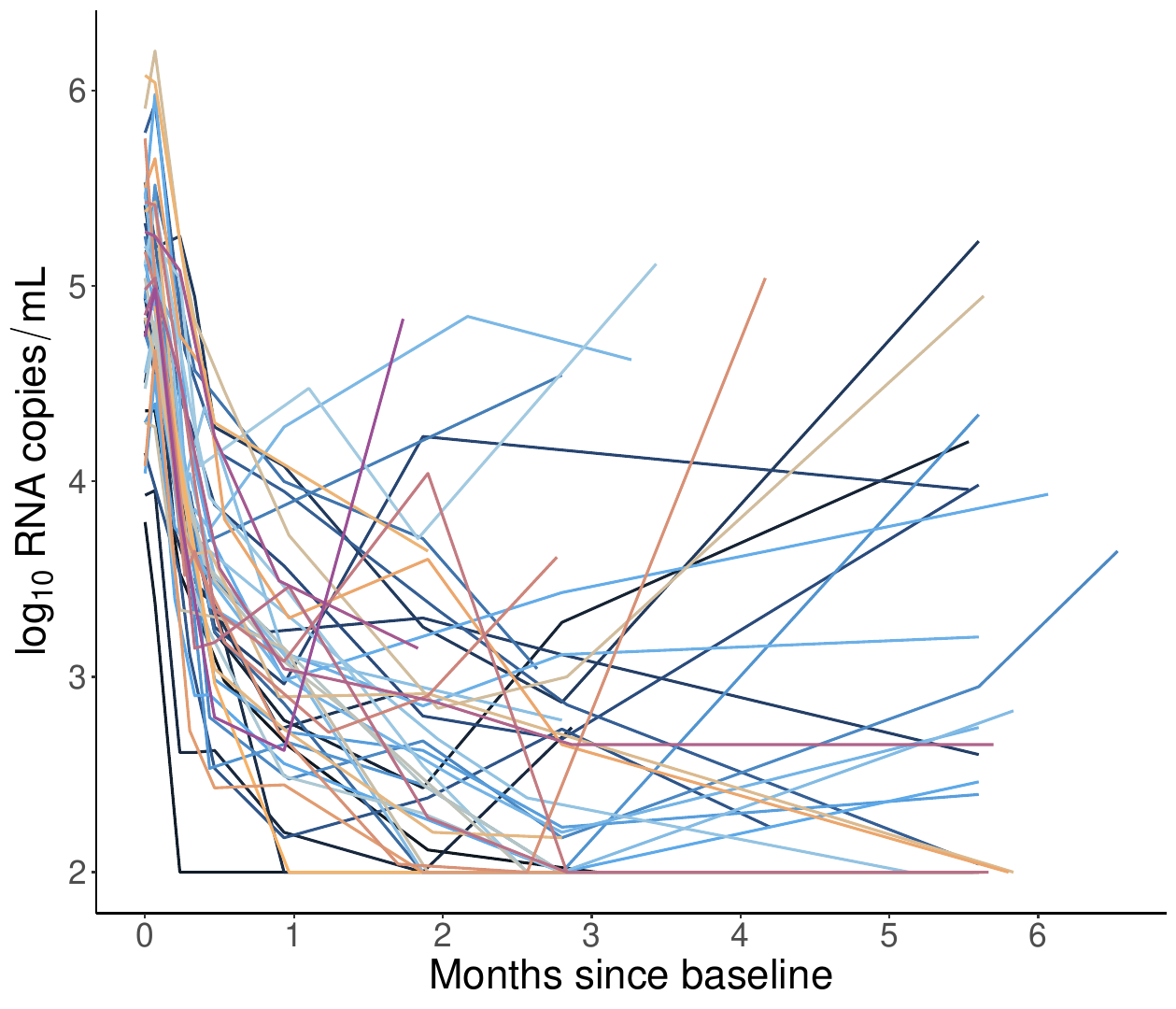}
      \caption{Longitudinal $\log_{10}$ viral load trajectories for 46 ACTG~315 participants. The censoring limit is 100~copies/mL ($\log_{10}=2$). Month~0 corresponds to the baseline measurement.}
      \label{fig:ACTG315_RNA}
    \end{figure}

    The profiles in \autoref{fig:ACTG315_RNA} show the familiar biphasic pattern of a rapid early decline after treatment initiation, followed by a much slower long-term phase. They also visually suggest between-subject heterogeneity in transition timing and possibly in the apparent width of the transition. To summarize timing heterogeneity descriptively, we fitted simple broken stick regressions separately for each participant over a grid of candidate changepoints and retained the value that minimized the residual sum of squares. For these broken stick fits, uncensored observations were used when at least 5 were available for a participant; otherwise, all available measurements were used, with values below the LLOQ set to the quantification limit. This yielded usable descriptive fits for 45 participants. The resulting changepoint estimates had a median of $0.60$ months, an interquartile range (IQR) of $0.40$ to $0.65$ months, and a range of $0.25$ to $2.45$ months. The corresponding median early slope was $-3.99$ $\log_{10}$ HIV-1 RNA (copies/mL) per month (IQR $-4.72$ to $-3.10$), whereas the median late slope was $-0.01$ (IQR $-0.39$ to $0.21$).
    
    Several late follow-up measurements deviate noticeably from the dominant trajectory, motivating a heavy-tailed residual specification. We therefore combine the smooth two-slope mean structure with a skew-$t$ residual density, consistent with previous work on heterogeneous and censored HIV viral load data, for example, Lachos et~al. \citeyearpar{lachos2011linear}. The residual model provides a robust likelihood for estimating the mean and heterogeneity parameters; the marginal quantities of interest are determined by the subject-specific mean structure and the fitted distributions of changepoints and transition sharpness.
    
    Earlier analyses of ACTG~315 include biphasic nonlinear mixed-effects models with exponential decay components \citep{wu2002joint,lachos2011linear}, as well as piecewise-linear mixed-effects models with one or more hard changepoints \citep{liang2007segmental,huang2013segmental}. Thus, ACTG~315 provides a familiar setting in which a two-phase structure is plausible, and the difference between plug-in and marginal summaries can be examined transparently.

    \section{Smooth random changepoint model} \label{sec:smooth-model}

    \subsection{Mean structure and smooth transition}
    
    We model subject-specific trajectories, allowing treatment-arm-specific parameters, as curves whose instantaneous slope moves from an early limiting value to a later limiting value. Let $j = 1,\ldots,J$ index treatment groups, and let $i = 1,\ldots,N_j$ index subjects assigned to treatment $j$. For each subject, let $k = 1,\ldots,n_{ij}$ index repeated measurements at times $t_{ijk}$. Time is measured from baseline, so $t = 0$ denotes the baseline visit, and the changepoint times are measured on the same time scale relative to baseline.

    To connect the early and late phases, we specify a smooth transition on the slope scale. For subject $i$ in arm $j$, we let
    \begin{equation}
        \frac{\mathrm{d}\mu_{ij}\left(t\right)}{\mathrm{d}t}
        =
        \beta_{1ij}
        +
        \beta_{2ij}
        \Phi\left(\kappa_{ij}\left(t - \tau_{ij}\right)\right),
        \label{eq:subject-slope}
    \end{equation}
    where $\Phi\left(\cdot\right)$ denotes the standard normal cumulative distribution function (CDF), $\beta_{1ij}$ is the early limiting slope, $\beta_{1ij} + \beta_{2ij}$ is the late limiting slope, $\tau_{ij}$ is the center of the transition, and $\kappa_{ij} > 0$ controls how quickly the transition occurs. The probit term $\Phi\left(\kappa_{ij}\left(t - \tau_{ij}\right)\right)$ acts as a smooth transition gate. When $t$ is well before $\tau_{ij}$, this gate is close to 0, and the instantaneous slope is close to $\beta_{1ij}$. When $t$ is well after $\tau_{ij}$, the gate is close to 1, and the instantaneous slope is close to $\beta_{1ij} + \beta_{2ij}$. The probit gate therefore gives a subject-specific transition that is continuous on the slope scale, rather than a hard change in slope at $\tau_{ij}$. At $t = \tau_{ij}$, the gate equals $1/2$, so
    \begin{equation}
        \left.\frac{\mathrm{d}\mu_{ij}\left(t\right)}{\mathrm{d}t}\right|_{t=\tau_{ij}}
        =
        \beta_{1ij}
        +
        \frac{1}{2}\beta_{2ij}.
    \end{equation}
    Thus, $\tau_{ij}$ is the time at which the subject has moved halfway, on the slope scale, from the early slope to the late slope.
    
    The parameter $\kappa_{ij}$ has an inverse-width interpretation. If $t_p$ denotes the time at which the transition gate equals $p$, then
    \begin{equation}
    t_p
    =
    \tau_{ij}
    +
    \frac{\Phi^{-1}\left(p\right)}{\kappa_{ij}}.
    \end{equation}
    Therefore, the time required for the slope gate to move from $10\%$ to $90\%$ of its transition is
    \begin{equation}
        w_{10,90,ij}
        =
        t_{0.90} - t_{0.10}
        =
        \frac{\Phi^{-1}\left(0.90\right) - \Phi^{-1}\left(0.10\right)}{\kappa_{ij}}
        \approx
        \frac{2.56}{\kappa_{ij}}.
    \label{eq:kappa-width}
    \end{equation}
    Thus, larger values of $\kappa_{ij}$ produce sharper transitions, whereas smaller values spread the transition over a longer interval. Allowing $\kappa_{ij}$ to vary permits the duration of the subject-level bend to differ between individuals.

    The corresponding mean trajectory is obtained by integrating \eqref{eq:subject-slope} from baseline. For compactness, define
    \begin{equation}
        h_{\kappa}\left(x\right)
        =
        x\Phi\left(\kappa x\right)
        +
        \frac{\phi\left(\kappa x\right)}{\kappa},
        \qquad \kappa > 0,
        \label{eq:softhinge}
    \end{equation}
    so that
    \begin{equation}
        \frac{\mathrm{d}}{\mathrm{d}x}h_{\kappa}\left(x\right)
        =
        \Phi\left(\kappa x\right).
    \end{equation}
    Equivalently, $h_{\kappa}\left(x\right)$ can be written as $\mathrm{E}\left[\left(x + Z/\kappa\right)^{+}\right]$ with $Z \sim \mathcal{N}\left(0,1\right)$. This representation gives a normal-convolution smoothing of the hard hinge $x^{+} = \max\left(x,0\right)$, related to the normal positive-part expectation in the Bachelier formulation \citep{bachelier1900theorie}; see also Cram{\'e}r \citeyearpar{cramer1946} for standard normal integral identities. As $\kappa \to \infty$, this representation approaches the abrupt hard hinge.
    
    Setting $\mu_{ij}\left(0\right) = \alpha_{0ij}$ and integrating from $0$ to $t$ gives
    \begin{equation}
        \begin{aligned}
            \mu_{ij}\left(t\right)
            &=
            \alpha_{0ij}
            +
            \int_{0}^{t}
            \left[
            \beta_{1ij}
            +
            \beta_{2ij}
            \Phi\left(\kappa_{ij}\left(u - \tau_{ij}\right)\right)
            \right]\mathrm{d}u \\
            &=
            \alpha_{0ij}
            +
            \beta_{1ij}t
            +
            \beta_{2ij}
            \left\{
            h_{\kappa_{ij}}\left(t - \tau_{ij}\right)
            -
            h_{\kappa_{ij}}\left(-\tau_{ij}\right)
            \right\}.
        \end{aligned}
        \label{eq:subjectcurve}
    \end{equation}
    The subtraction of $h_{\kappa_{ij}}\left(-\tau_{ij}\right)$ anchors the trajectory at baseline, so that $\mu_{ij}\left(0\right) = \alpha_{0ij}$. The mean trajectory at the observed measurement times is then
    \begin{equation}
        \mu_{ijk}
        =
        \mu_{ij}\left(t_{ijk}\right),
        \label{eq:mean}
    \end{equation}
    for $k = 1,\ldots,n_{ij}$.

\begin{figure}[!htbp]
        \centering
    
        \begin{subfigure}{0.48\textwidth}
            \centering
            \includegraphics[width=\linewidth]{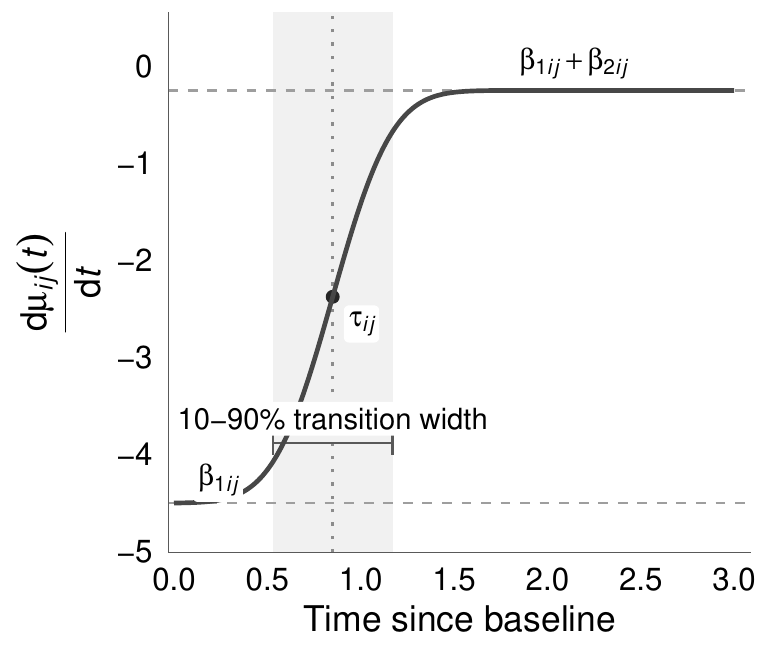}
            \caption{One subject: instantaneous slope}
            \label{fig:subject-reference-slope}
        \end{subfigure}
        \hfill
        \begin{subfigure}{0.48\textwidth}
            \centering
            \includegraphics[width=\linewidth]{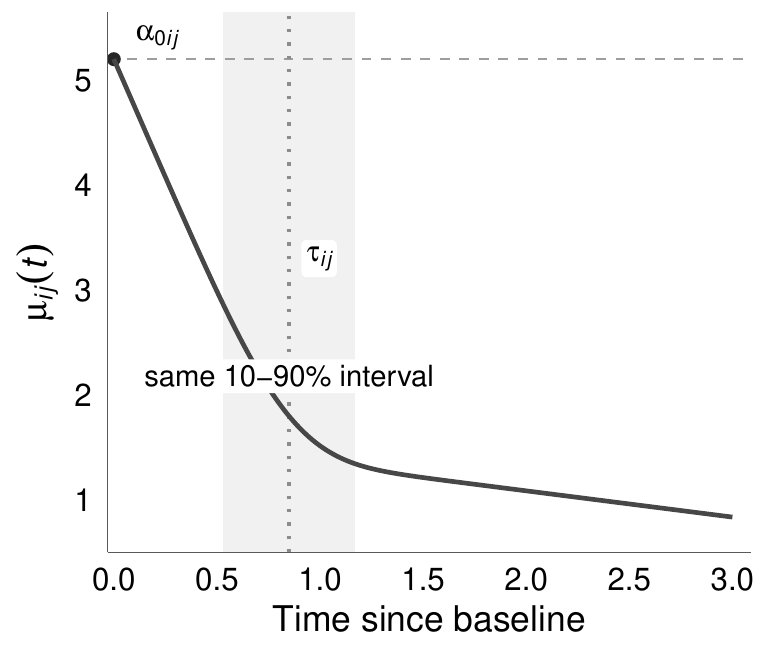}
            \caption{Same subject: mean trajectory}
            \label{fig:subject-reference-mean}
        \end{subfigure}
    
        \vspace{0.8em}
    
        \begin{subfigure}{0.48\textwidth}
            \centering
            \includegraphics[width=\linewidth]{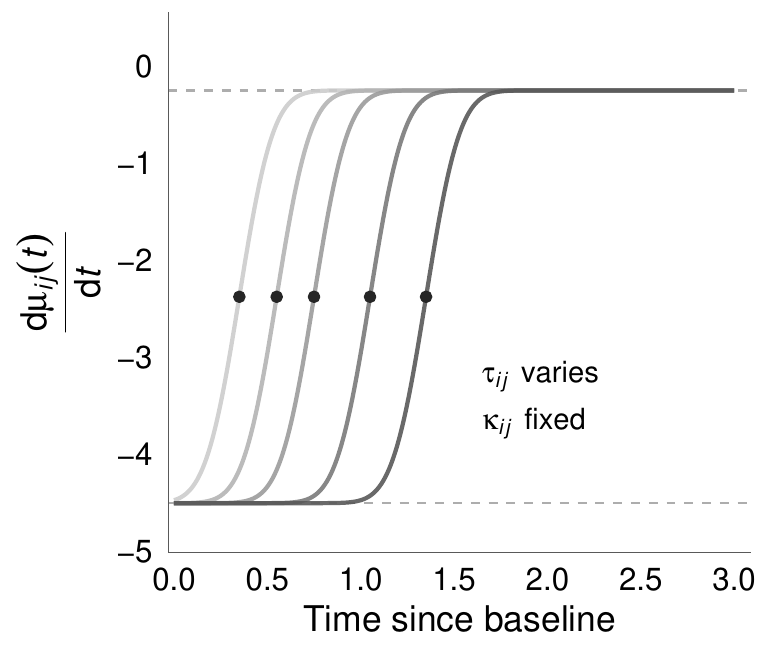}
            \caption{Several subjects: varying $\tau_{ij}$ only}
            \label{fig:subject-random-tau}
        \end{subfigure}
        \hfill
        \begin{subfigure}{0.48\textwidth}
            \centering
            \includegraphics[width=\linewidth]{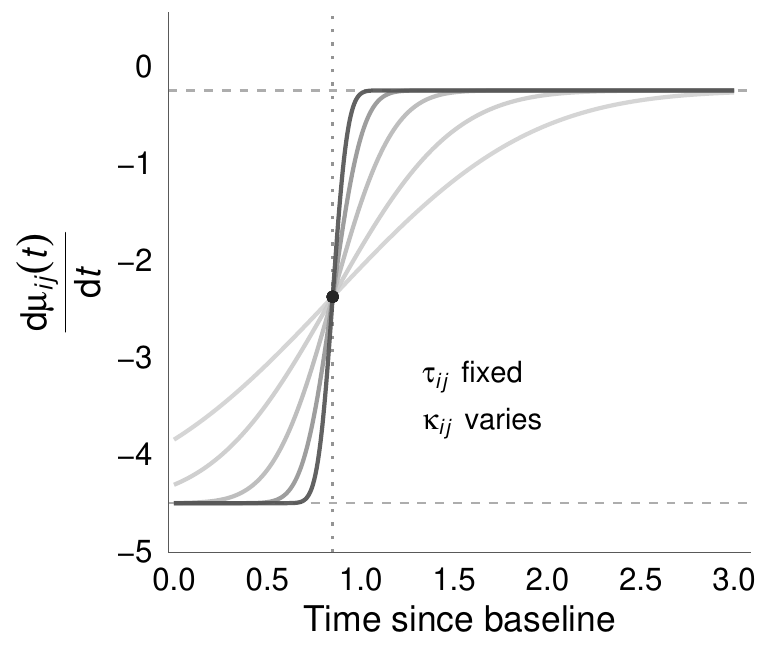}
            \caption{Several subjects: varying $\kappa_{ij}$ only}
            \label{fig:subject-random-kappa}
        \end{subfigure}
    
        \caption{Subject-level interpretation of the smooth changepoint model. Panel~(a) shows the instantaneous slope for one subject. The probit gate moves the slope from the early limiting value $\beta_{1ij}$ to the late limiting value $\beta_{1ij} + \beta_{2ij}$, with $\tau_{ij}$ marking the midpoint of the transition and the indicated interval marking the approximate $10\%$ to $90\%$ transition width. Panel~(b) shows the corresponding subject-specific mean trajectory obtained by integrating the slope from baseline, with the same transition interval shown for reference. Panel~(c) illustrates heterogeneity in changepoint timing by varying $\tau_{ij}$ while holding $\kappa_{ij}$ fixed. Panel~(d) illustrates heterogeneity in transition width by varying $\kappa_{ij}$ while holding $\tau_{ij}$ fixed. The population-average summaries are obtained by averaging over these subject-level components.}
        \label{fig:subject-level-schematic}
    \end{figure}

    The subject-level interpretation of the slope, baseline-anchored mean trajectory, changepoint timing, and transition width is summarized in \autoref{fig:subject-level-schematic}.

    This baseline-anchored form is the convention used throughout the marginal summaries below: mean kernels are defined for the centered contribution $h_{\kappa_{ij}}\left(t - \tau_{ij}\right) - h_{\kappa_{ij}}\left(-\tau_{ij}\right)$.
    
    The abrupt two-slope model is recovered in the limit as $\kappa_{ij} \to \infty$, which replaces the smooth hinge by the hard hinge. A model with common transition sharpness is obtained by setting $\kappa_{ij} = \kappa_j$ for all subjects in arm $j$.

    \subsection{Between-subject heterogeneity}
    
    Between-subject heterogeneity is modeled through random effects on the baseline intercept and the two slope parameters, together with arm-specific distributions for changepoint timing and transition sharpness. Specifically,
    \begin{equation}
      \begin{bmatrix}
        \alpha_{0ij}\\[2pt]\beta_{1ij}\\[2pt]\beta_{2ij}
      \end{bmatrix}
      \sim
      \mathcal{N}_{3}\left(
      \begin{bmatrix}
        \alpha_{0j}\\[2pt]\beta_{1j}\\[2pt]\beta_{2j}
      \end{bmatrix},
      \boldsymbol{\Sigma}_{v,j}
      \right),
    \label{eq:re}
    \end{equation}
    where $\alpha_{0j}$, $\beta_{1j}$, and $\beta_{2j}$ are the arm-level means of the corresponding subject-specific parameters, and $\boldsymbol{\Sigma}_{v,j}$ is an unstructured $3 \times 3$ covariance matrix for treatment arm $j$. We also assume
    \begin{equation}
      \tau_{ij}
      \sim
      \mathrm{TN}\left(\mu_{\tau,j}, \sigma_{\tau,j}^{2}; 0, \tau_{\max}\right),
      \qquad
      \kappa_{ij}
      \sim
      \mathrm{TN}\left(\mu_{\kappa,j}, \sigma_{\kappa,j}^{2}; 0, \infty\right),
    \label{eq:tau-kappa}
    \end{equation}
    where $\mathrm{TN}\left(\mu,\sigma^{2};a,b\right)$ denotes a normal distribution with pre-truncation location $\mu$ and variance $\sigma^{2}$, truncated to $\left[a,b\right]$, and $\tau_{\max}$ is a prespecified upper bound for changepoint timing on the analysis time scale. The distribution of $\tau_{ij}$ describes variation in transition timing, while the distribution of $\kappa_{ij}$ describes variation in subject-level transition sharpness.
    
    The estimands below do not require independence, but the deterministic formulas and the ACTG implementation use a working independence structure in which $\left(\alpha_{0ij},\beta_{1ij},\beta_{2ij}\right)$ is independent of $\left(\tau_{ij},\kappa_{ij}\right)$ and $\tau_{ij}$ is independent of $\kappa_{ij}$, conditionally on the corresponding arm-level parameters. Correlation among $\left(\alpha_{0ij},\beta_{1ij},\beta_{2ij}\right)$ is still captured through $\boldsymbol{\Sigma}_{v,j}$. This independence structure simplifies the closed-form evaluation of the marginal mean and slope, but it is not part of the definition of the population-average estimands. If dependence is expected, for example if subjects with steeper early declines tend to bend earlier or more sharply, the same estimands can be evaluated by Monte Carlo averaging under the fitted joint distribution of $\left(\alpha_{0ij},\beta_{1ij},\beta_{2ij},\tau_{ij},\kappa_{ij}\right)$.

    \subsection{Residual model and likelihood}
    
    For notational simplicity, we first describe the uncensored formulation. Conditional on the subject-specific mean trajectory, the outcome follows an additive error model:
    \begin{equation}
        Y_{ijk}
        =
        \mu_{ijk}
        +
        \varepsilon_{ijk},
        \qquad
        \varepsilon_{ijk}
        \sim
        f_{\varepsilon}\left(\cdot; \boldsymbol{\theta}_j\right),
        \qquad
        \mathrm{E}\left(\varepsilon_{ijk}\right) = 0.
        \label{eq:resid}
    \end{equation}
    where $\boldsymbol{\theta}_j$ collects dispersion, skewness, and tail thickness parameters for treatment arm $j$. Examples include Gaussian, Student-$t$, and skew-$t$ residual families, parameterized so that $\mathrm{E}\left(\varepsilon_{ijk}\right) = 0$. Heavy-tailed residual specifications are a natural extension of nonlinear mixed-effects models with non-Gaussian residual variation; see, for example, the general discussion in Davidian and Giltinan \citeyearpar{davidian1995nonlinear,davidian2003nonlinear}. The residual model affects inference for the mean structure and heterogeneity parameters but does not change the definitions of the arm-level marginal mean and slope, which are determined by the subject-specific mean trajectory and the assumed hierarchical model for its subject-level parameters. In the Bayesian implementation for ACTG~315, we take $f_{\varepsilon}$ to be the Hansen skew-$t$ density \citep{hansen1994autoregressive}.
    
    We assume that, conditional on the subject-specific parameters $\left(\alpha_{0ij}, \beta_{1ij}, \beta_{2ij}, \tau_{ij}, \kappa_{ij}\right)$ and the arm-level residual parameters $\boldsymbol{\theta}_j$, the repeated measurements $Y_{ijk}$ for $k = 1,\ldots,n_{ij}$ are independent. Under the uncensored formulation, the subject-level likelihood is
    \begin{equation}
        \mathcal{L}_{ij}
        =
        \prod_{k=1}^{n_{ij}}
        f_{\varepsilon}\left(
            Y_{ijk}
            -
            \mu_{ij}\left(t_{ijk}\right);
            \boldsymbol{\theta}_j
        \right),
        \label{eq:lik-subject}
    \end{equation}
    where $f_{\varepsilon}$ is the density of the residual error in \eqref{eq:resid}. Accordingly, the conditional data likelihood factors over subjects and treatment arms:
    \begin{equation}
        \mathcal{L}
        =
        \prod_{j=1}^{J}
        \prod_{i=1}^{N_j}
        \mathcal{L}_{ij}.
        \label{eq:lik-full}
    \end{equation}

    \section{Population-average mean trajectory and slope}
    \label{sec:marginal-trajectories}
    
    This section defines the treatment-arm summaries used in the rest of the paper. The target is the model-implied population-average trajectory for a randomly selected subject from treatment arm $j$, together with the corresponding population-average instantaneous slope. These quantities are not empirical averages of the fitted curves for the observed participants, nor are they the curve for a hypothetical subject with all subject-specific parameters fixed at typical values.
    
    Here and below, $j$ indexes the treatment arm and $i$ denotes a generic subject drawn from that arm. Let
    \begin{equation}
        \boldsymbol{\psi}_{ij}
        =
        \left(
        \alpha_{0ij},
        \beta_{1ij},
        \beta_{2ij},
        \tau_{ij},
        \kappa_{ij}
        \right)^{\top}
    \end{equation}
    collect the subject-level parameters that determine the underlying trajectory for a randomly selected subject from arm $j$. Let $\boldsymbol{\Omega}_j$ collect the corresponding arm-level parameters that define the distribution of $\boldsymbol{\psi}_{ij}$. We write
    \begin{equation}
        \mathrm{E}_{j}\left[\cdot\right]
        =
        \mathrm{E}\left[\cdot \middle| \boldsymbol{\Omega}_j\right]
    \end{equation}
    for expectation over the arm-specific distribution of the subject-level parameters. This expectation is over between-subject heterogeneity in baseline level, slopes, changepoint timing, and transition sharpness, not over residual error. In the Bayesian implementation, this conditional expectation is evaluated for each retained posterior draw of $\boldsymbol{\Omega}_j$, and posterior uncertainty is then summarized across draws.
    
    The population-average mean trajectory is
    \begin{equation}
        \mu_{\mathrm{marg},j}\left(t\right)
        =
        \mathrm{E}_{j}\left[
        \mu_{ij}\left(t\right)
        \right].
        \label{eq:mu-marg-def}
    \end{equation}
    The corresponding population-average instantaneous slope is obtained by averaging the subject-specific instantaneous slopes:
    \begin{equation}
        s_{\mathrm{marg},j}\left(t\right)
        =
        \mathrm{E}_{j}\left[
        \frac{\mathrm{d}}{\mathrm{d}t}\mu_{ij}\left(t\right)
        \right]
        =
        \mathrm{E}_{j}\left[
        \beta_{1ij}
        +
        \beta_{2ij}
        \Phi\left(\kappa_{ij}\left(t - \tau_{ij}\right)\right)
        \right].
        \label{eq:slope-marg-def}
    \end{equation}
    Because the smooth transition gate lies between 0 and 1, the subject-specific slope is bounded in magnitude by $\left|\beta_{1ij}\right| + \left|\beta_{2ij}\right|$. Under the finite first-moment assumptions of the mixed model, the dominated convergence theorem justifies interchanging differentiation and expectation \citep{Billingsley1995,Folland1999}, so $s_{\mathrm{marg},j}\left(t\right)$ is also the derivative of $\mu_{\mathrm{marg},j}\left(t\right)$ with respect to $t$. In the censored ACTG application, $\mu_{\mathrm{marg},j}\left(t\right)$ is the marginal mean of the underlying latent viral load process, not the mean of the recorded values after applying the LLOQ. The residual distribution affects parameter estimation, but because it is centered at zero, it does not alter the definitions in \eqref{eq:mu-marg-def} and \eqref{eq:slope-marg-def}.

    Substituting the subject-specific trajectory from \eqref{eq:subjectcurve} gives the fully marginal form
    \begin{equation}
        \mu_{\mathrm{marg},j}\left(t\right)
        =
        \mathrm{E}_{j}\left(\alpha_{0ij}\right)
        +
        \mathrm{E}_{j}\left(\beta_{1ij}\right)t
        +
        \mathrm{E}_{j}\left[
        \beta_{2ij}
        \left\{
        h_{\kappa_{ij}}\left(t - \tau_{ij}\right)
        -
        h_{\kappa_{ij}}\left(-\tau_{ij}\right)
        \right\}
        \right],
        \label{eq:mu-marg-general}
    \end{equation}
    and
    \begin{equation}
        s_{\mathrm{marg},j}\left(t\right)
        =
        \mathrm{E}_{j}\left(\beta_{1ij}\right)
        +
        \mathrm{E}_{j}\left[
        \beta_{2ij}
        \Phi\left(\kappa_{ij}\left(t - \tau_{ij}\right)\right)
        \right].
        \label{eq:slope-marg-general}
    \end{equation}
    Equations \eqref{eq:mu-marg-general} and \eqref{eq:slope-marg-general} are the estimands. They remain valid if the level, slope, changepoint, and transition sharpness parameters are dependent. In that more general case, the expectations can be evaluated directly by Monte Carlo averaging under the fitted joint distribution. The deterministic simplifications below use the independence structure specified in \autoref{sec:smooth-model}: $\left(\alpha_{0ij},\beta_{1ij},\beta_{2ij}\right)$ is independent of $\left(\tau_{ij},\kappa_{ij}\right)$, and $\tau_{ij}$ is independent of $\kappa_{ij}$.

    For the kernel calculations, let $\tilde{\tau}_j$ and $\tilde{\kappa}_j$ denote generic random variables with the same fitted distributions as $\tau_{ij}$ and $\kappa_{ij}$, respectively, for a randomly selected subject in arm $j$. Thus, $\tilde{\tau}_j \overset{d}{=} \tau_{ij}$ and $\tilde{\kappa}_j \overset{d}{=} \kappa_{ij}$.

    Equation \eqref{eq:re} gives $\mathrm{E}_{j}\left(\alpha_{0ij}\right) = \alpha_{0j}$, $\mathrm{E}_{j}\left(\beta_{1ij}\right) = \beta_{1j}$, and $\mathrm{E}_{j}\left(\beta_{2ij}\right) = \beta_{2j}$. Together with the independence assumptions above, this permits the remaining expectation to be evaluated by first averaging over changepoint timing and then averaging over transition sharpness. For a fixed value $\kappa > 0$, define the baseline-centered changepoint-marginalized hinge kernel
    \begin{equation}
        H_j\left(t \middle| \kappa\right)
        =
        \mathrm{E}_{\tilde{\tau}_j}\left[
        h_{\kappa}\left(t - \tilde{\tau}_j\right)
        -
        h_{\kappa}\left(-\tilde{\tau}_j\right)
        \right],
        \label{eq:Hj-def-pop}
    \end{equation}
    and the corresponding changepoint-marginalized slope kernel
    \begin{equation}
        G_j\left(t \middle| \kappa\right)
        =
        \mathrm{E}_{\tilde{\tau}_j}\left[
        \Phi\left(\kappa\left(t - \tilde{\tau}_j\right)\right)
        \right].
        \label{eq:Gj-def-pop}
    \end{equation}
    Averaging these kernels over the arm-specific distribution of $\tilde{\kappa}_j$ gives the corresponding fully marginalized kernels
    \begin{equation}
        \begin{aligned}
            \bar{H}_j\left(t\right)
            &=
            \mathrm{E}_{\tilde{\kappa}_j}\left[
            H_j\left(t \middle| \tilde{\kappa}_j\right)
            \right] \\
            &=
            \mathrm{E}_{\tilde{\kappa}_j}\left[
            \mathrm{E}_{\tilde{\tau}_j}\left[
            h_{\tilde{\kappa}_j}\left(t - \tilde{\tau}_j\right)
            -
            h_{\tilde{\kappa}_j}\left(-\tilde{\tau}_j\right)
            \right]
            \right],
        \end{aligned}
        \label{eq:Hbar-def}
    \end{equation}
    and
    \begin{equation}
        \begin{aligned}
            \bar{G}_j\left(t\right)
            &=
            \mathrm{E}_{\tilde{\kappa}_j}\left[
            G_j\left(t \middle| \tilde{\kappa}_j\right)
            \right] \\
            &=
            \mathrm{E}_{\tilde{\kappa}_j}\left[
            \mathrm{E}_{\tilde{\tau}_j}\left[
            \Phi\left(\tilde{\kappa}_j\left(t - \tilde{\tau}_j\right)\right)
            \right]
            \right].
        \end{aligned}
        \label{eq:Gbar-def}
    \end{equation}
    Under these assumptions, the population-average trajectory and slope reduce to
    \begin{equation}
        \mu_{\mathrm{marg},j}\left(t\right)
        =
        \alpha_{0j}
        +
        \beta_{1j}t
        +
        \beta_{2j}\bar{H}_j\left(t\right),
        \label{eq:mu-marg-separated}
    \end{equation}
    and
    \begin{equation}
        s_{\mathrm{marg},j}\left(t\right)
        =
        \beta_{1j}
        +
        \beta_{2j}\bar{G}_j\left(t\right).
        \label{eq:slope-marg-separated}
    \end{equation}
    Because $H_j\left(0 \middle| \kappa\right) = 0$ by construction, $\bar{H}_j\left(0\right) = 0$ and hence $\mu_{\mathrm{marg},j}\left(0\right) = \alpha_{0j}$. Also, since $G_j\left(t \middle| \kappa\right)$ is the derivative of $H_j\left(t \middle| \kappa\right)$ with respect to $t$,
    \begin{equation}
        \frac{\mathrm{d}}{\mathrm{d}t}\bar{H}_j\left(t\right)
        =
        \bar{G}_j\left(t\right),
        \label{eq:Hbar-deriv}
    \end{equation}
    and, since the transition gate is bounded, Fubini's theorem permits the expectation and the baseline-to-time integral to be interchanged \citep{Folland1999}. The marginal mean can therefore equivalently be written as
    \begin{equation}
        \mu_{\mathrm{marg},j}\left(t\right)
        =
        \alpha_{0j}
        +
        \beta_{1j}t
        +
        \beta_{2j}
        \int_{0}^{t}\bar{G}_j\left(u\right)\mathrm{d}u.
        \label{eq:mu-marg-integrated}
    \end{equation}
    For comparison, the plug-in trajectory fixes the changepoint and transition sharpness at the means of their fitted arm-level truncated-normal distributions. Let
    \begin{equation}
        \bar{\tau}_j
        =
        \mathrm{E}_{\tilde{\tau}_j}\left(\tilde{\tau}_j\right),
        \qquad
        \bar{\kappa}_j
        =
        \mathrm{E}_{\tilde{\kappa}_j}\left(\tilde{\kappa}_j\right),
        \label{eq:tau-kappa-trunc-means}
    \end{equation}
    where these expectations are taken under the corresponding truncated normal distributions, not under the untruncated normal laws. The plug-in trajectory is
    \begin{equation}
        \mu_{\mathrm{cond},j}\left(t\right)
        =
        \alpha_{0j}
        +
        \beta_{1j}t
        +
        \beta_{2j}
        \left\{
        h_{\bar{\kappa}_j}\left(t - \bar{\tau}_j\right)
        -
        h_{\bar{\kappa}_j}\left(-\bar{\tau}_j\right)
        \right\},
        \label{eq:mu-plugin}
    \end{equation}
    with plug-in slope
    \begin{equation}
        s_{\mathrm{cond},j}\left(t\right)
        =
        \beta_{1j}
        +
        \beta_{2j}
        \Phi\left(\bar{\kappa}_j\left(t - \bar{\tau}_j\right)\right).
        \label{eq:slope-plugin}
    \end{equation}
    These plug-in summaries describe a hypothetical representative subject. They generally differ from \eqref{eq:mu-marg-separated} and \eqref{eq:slope-marg-separated} because evaluating the nonlinear transition function at $\bar{\tau}_j$ and $\bar{\kappa}_j$ is not equivalent to averaging it over the distributions of $\tilde{\tau}_j$ and $\tilde{\kappa}_j$.

    \subsection{Evaluation of the changepoint-marginalized kernels}
    \label{sec:cond-tau}

    We first evaluate the inner expectation over the changepoint distribution while treating $\kappa$ as fixed. Suppose
    \begin{equation}
        \tilde{\tau}_j
        \sim
        \mathrm{TN}\left(\mu_{\tau,j}, \sigma_{\tau,j}^{2}; L_{\tau,j}, R_{\tau,j}\right).
        \label{eq:tau-tn-section4}
    \end{equation}
    In the ACTG application, we set $L_{\tau,j} = 0$ and $R_{\tau,j} = \tau_{\max}$, so changepoints are constrained to lie within the prespecified interval for changepoint timing.

    Define
    \begin{equation}
        z_{L,j}
        =
        \frac{L_{\tau,j} - \mu_{\tau,j}}{\sigma_{\tau,j}},
        \qquad
        z_{U,j}
        =
        \frac{R_{\tau,j} - \mu_{\tau,j}}{\sigma_{\tau,j}},
        \qquad
        Z_{\tau,j}
        =
        \Phi\left(z_{U,j}\right)
        -
        \Phi\left(z_{L,j}\right).
        \label{eq:trunc-normal-z}
    \end{equation}
    For compactness, also define
    \begin{equation}
        m_j\left(t\right)
        =
        t - \mu_{\tau,j},
        \qquad
        a_j\left(\kappa\right)
        =
        \kappa\sigma_{\tau,j},
        \qquad
        b_j\left(t,\kappa\right)
        =
        \kappa m_j\left(t\right),
        \qquad
        d_j\left(\kappa\right)
        =
        \sqrt{1 + a_j\left(\kappa\right)^2}.
        \label{eq:kernel-abd}
    \end{equation}
    Then the changepoint-marginalized slope kernel has the closed form
    \begin{equation}
        G_j\left(t \middle| \kappa\right)
        =
        \frac{
        \Phi_{2}\left(
        z_{U,j},
        \frac{b_j\left(t,\kappa\right)}{d_j\left(\kappa\right)};
        \frac{a_j\left(\kappa\right)}{d_j\left(\kappa\right)}
        \right)
        -
        \Phi_{2}\left(
        z_{L,j},
        \frac{b_j\left(t,\kappa\right)}{d_j\left(\kappa\right)};
        \frac{a_j\left(\kappa\right)}{d_j\left(\kappa\right)}
        \right)
        }{
        Z_{\tau,j}
        },
        \label{eq:G-trunc-i}
    \end{equation}
    where $\Phi_2\left(x,y;\rho\right)$ denotes the bivariate standard normal CDF with correlation $\rho$. The corresponding closed-form calculation for the baseline-centered mean kernel $H_j\left(t \middle| \kappa\right)$ is given in \autoref{app:Etau-derivation}.

    The untruncated normal case gives the clearest interpretation. If $\left(L_{\tau,j},R_{\tau,j}\right) = \left(-\infty,\infty\right)$, define the effective transition width
    \begin{equation}
        w_j\left(\kappa\right)
        =
        \sqrt{\sigma_{\tau,j}^{2} + \frac{1}{\kappa^2}}.
        \label{eq:effective-width}
    \end{equation}
    Then the baseline-centered mean kernel is
    \begin{equation}
    \begin{aligned}
        H_j\left(t \middle| \kappa\right)
        &=
        m_j\left(t\right)
        \Phi\left(
        \frac{m_j\left(t\right)}{w_j\left(\kappa\right)}
        \right)
        +
        w_j\left(\kappa\right)
        \phi\left(
        \frac{m_j\left(t\right)}{w_j\left(\kappa\right)}
        \right)
        \\
        &\quad
        -
        m_j\left(0\right)
        \Phi\left(
        \frac{m_j\left(0\right)}{w_j\left(\kappa\right)}
        \right)
        -
        w_j\left(\kappa\right)
        \phi\left(
        \frac{m_j\left(0\right)}{w_j\left(\kappa\right)}
        \right).
    \end{aligned}
        \label{eq:Etau-untrunc-i}
    \end{equation}
    and the slope kernel is
    \begin{equation}
        G_j\left(t \middle| \kappa\right)
        =
        \Phi\left(
        \frac{m_j\left(t\right)}{w_j\left(\kappa\right)}
        \right).
        \label{eq:G-untrunc-i}
    \end{equation}
    Thus, in the untruncated case, variation in changepoint timing and subject-level smoothness combine through $w_j\left(\kappa\right)$. The term $\sigma_{\tau,j}$ broadens the population transition because subjects bend at different times, while $1/\kappa$ broadens it because each individual bend is gradual.

    \subsection{Averaging over transition sharpness}
    \label{sec:marg-kappa}

    The remaining step is to average over
    \begin{equation}
        \tilde{\kappa}_j
        \sim
        \mathrm{TN}\left(\mu_{\kappa,j}, \sigma_{\kappa,j}^{2}; 0, \infty\right).
        \label{eq:kappa-tn-section4}
    \end{equation}
    This leaves a one-dimensional numerical integration because the inner expectation over $\tilde{\tau}_j$ has already been evaluated analytically. Let $F_{\kappa,j}$ denote the CDF of the distribution of $\tilde{\kappa}_j$ in \eqref{eq:kappa-tn-section4}.

    The baseline-centered definition of $H_j\left(t \middle| \kappa\right)$ is important for this step. As $\kappa \downarrow 0$,
    \begin{equation}
        h_{\kappa}\left(x\right)
        =
        \frac{\phi\left(0\right)}{\kappa}
        +
        \frac{x}{2}
        +
        O\left(\kappa\right).
        \label{eq:h-small-kappa}
    \end{equation}
    The common divergent $1/\kappa$ term is removed by the baseline-centering subtraction in \eqref{eq:Hj-def-pop}. Thus, $\bar{H}_j\left(t\right)$ can be evaluated by averaging $H_j\left(t \middle| \kappa\right)$ over the fitted distribution of $\tilde{\kappa}_j$. Equivalently, it can be obtained by integrating the bounded function $\bar{G}_j\left(t\right)$ from baseline, as in \eqref{eq:mu-marg-integrated}.

    For a deterministic midpoint quadrature with $K$ equal-probability points, set
    \begin{equation}
        q_r
        =
        \frac{r - \frac{1}{2}}{K},
        \qquad
        \kappa_{\left(r\right)j}
        =
        F_{\kappa,j}^{-1}\left(q_r\right),
        \qquad
        r = 1,\ldots,K.
        \label{eq:kappa-quad-points}
    \end{equation}
    Then
    \begin{equation}
        \bar{H}_j\left(t\right)
        \approx
        \frac{1}{K}
        \sum_{r=1}^{K}
        H_j\left(t \middle| \kappa_{\left(r\right)j}\right),
        \label{eq:Hbar-quad}
    \end{equation}
    and
    \begin{equation}
        \bar{G}_j\left(t\right)
        \approx
        \frac{1}{K}
        \sum_{r=1}^{K}
        G_j\left(t \middle| \kappa_{\left(r\right)j}\right).
        \label{eq:Gbar-quad}
    \end{equation}
    In the analyses below, we use this deterministic quadrature for every posterior draw. It avoids Monte Carlo noise in the reported marginal trajectories and slopes, while still averaging over the fitted transition sharpness distribution.

\begin{figure}[!htbp]
        \centering
    
        \begin{subfigure}{0.48\textwidth}
            \centering
            \includegraphics[width=\linewidth]{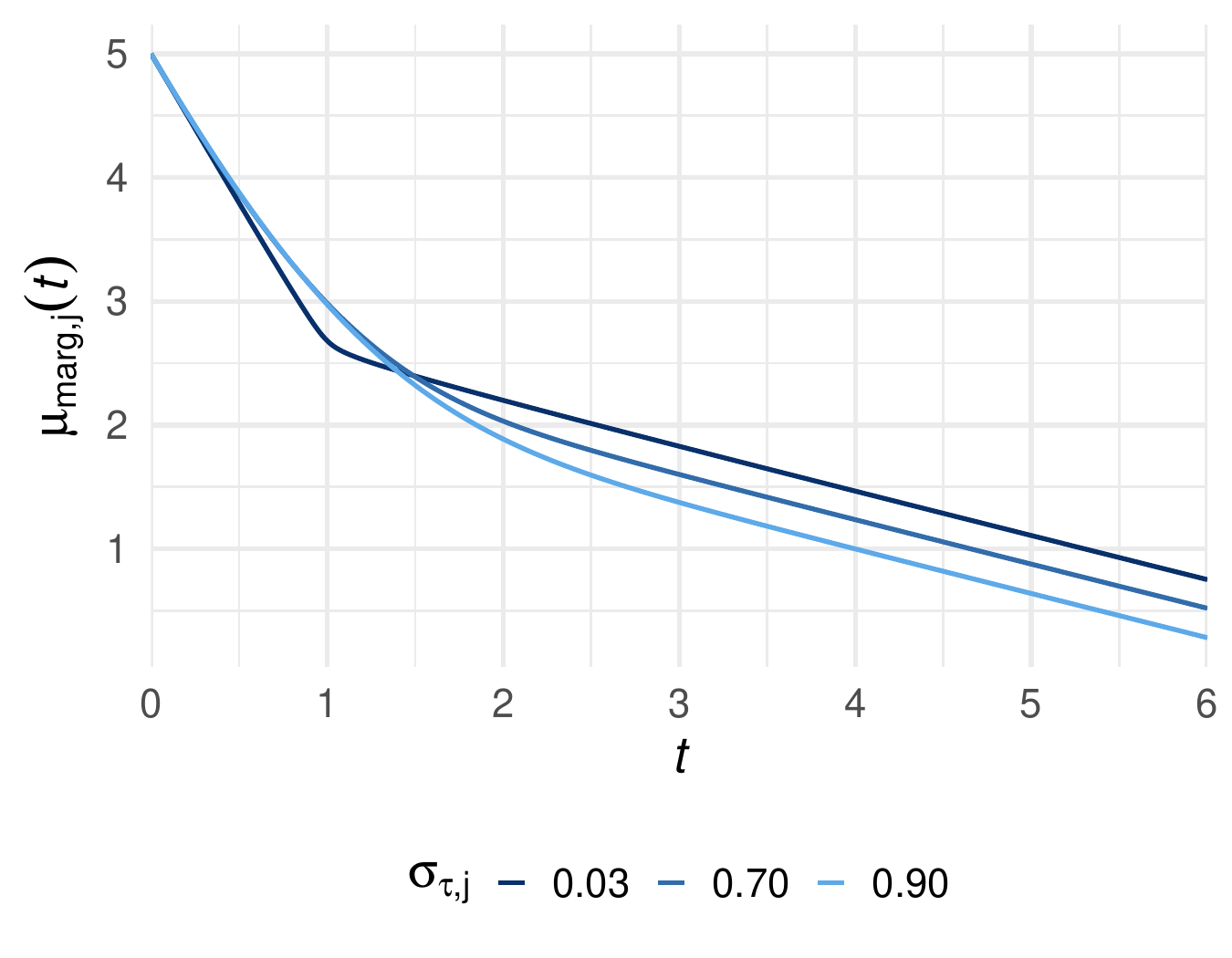}
    
            \vspace{0.6em}
    
            \includegraphics[width=\linewidth]{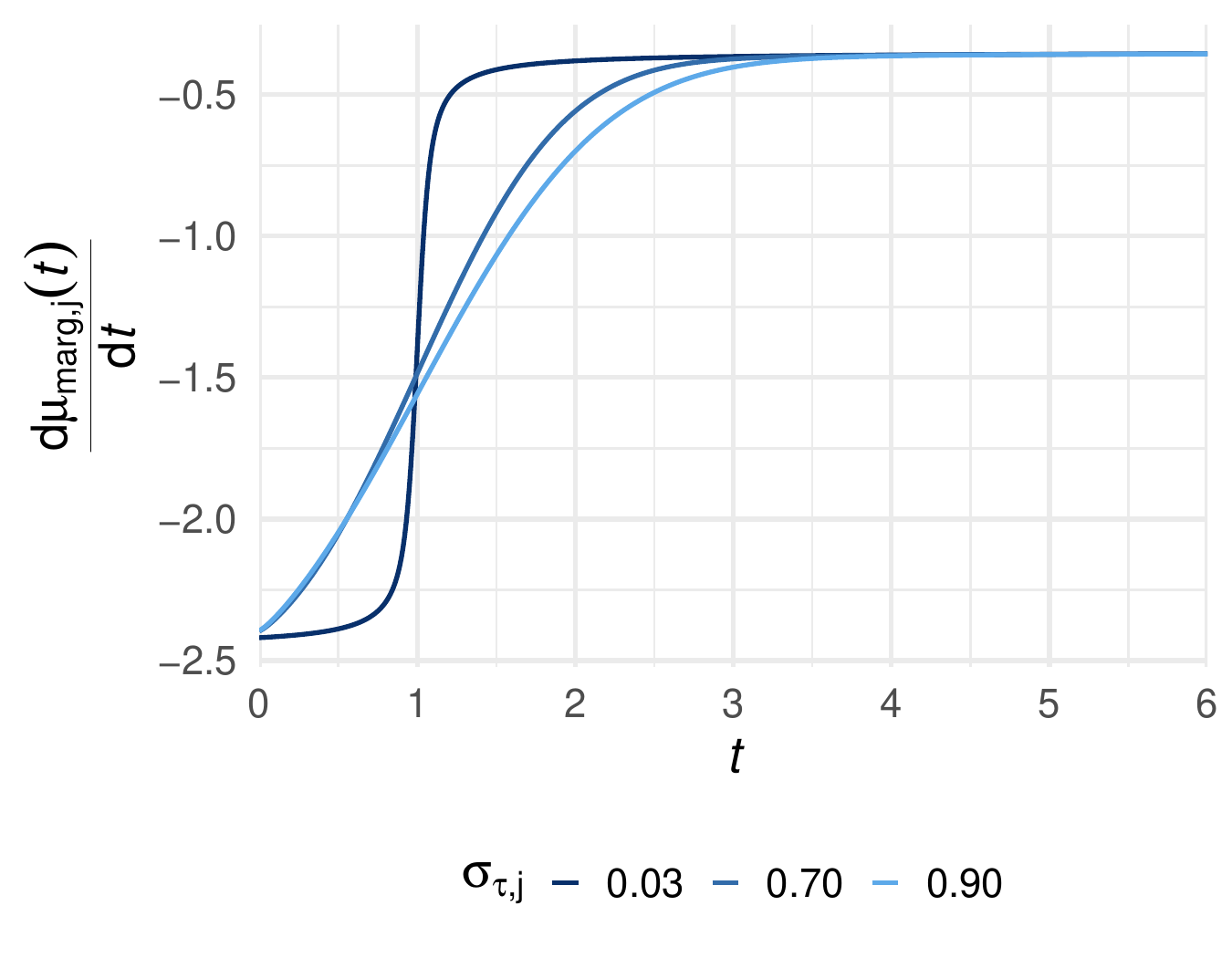}
            \caption{Varying $\sigma_{\tau}$. Top: mean trajectory. Bottom: instantaneous slope. Fixed $\kappa_{ij} \sim \mathrm{TN}\left(2, 20^{2}; 0, \infty\right)$ with $\sigma_{\tau,j} \in \left\{0.03, 0.70, 0.90\right\}$.}
            \label{fig:slope-sigmat}
        \end{subfigure}
        \hfill
        \begin{subfigure}{0.48\textwidth}
            \centering
            \includegraphics[width=\linewidth]{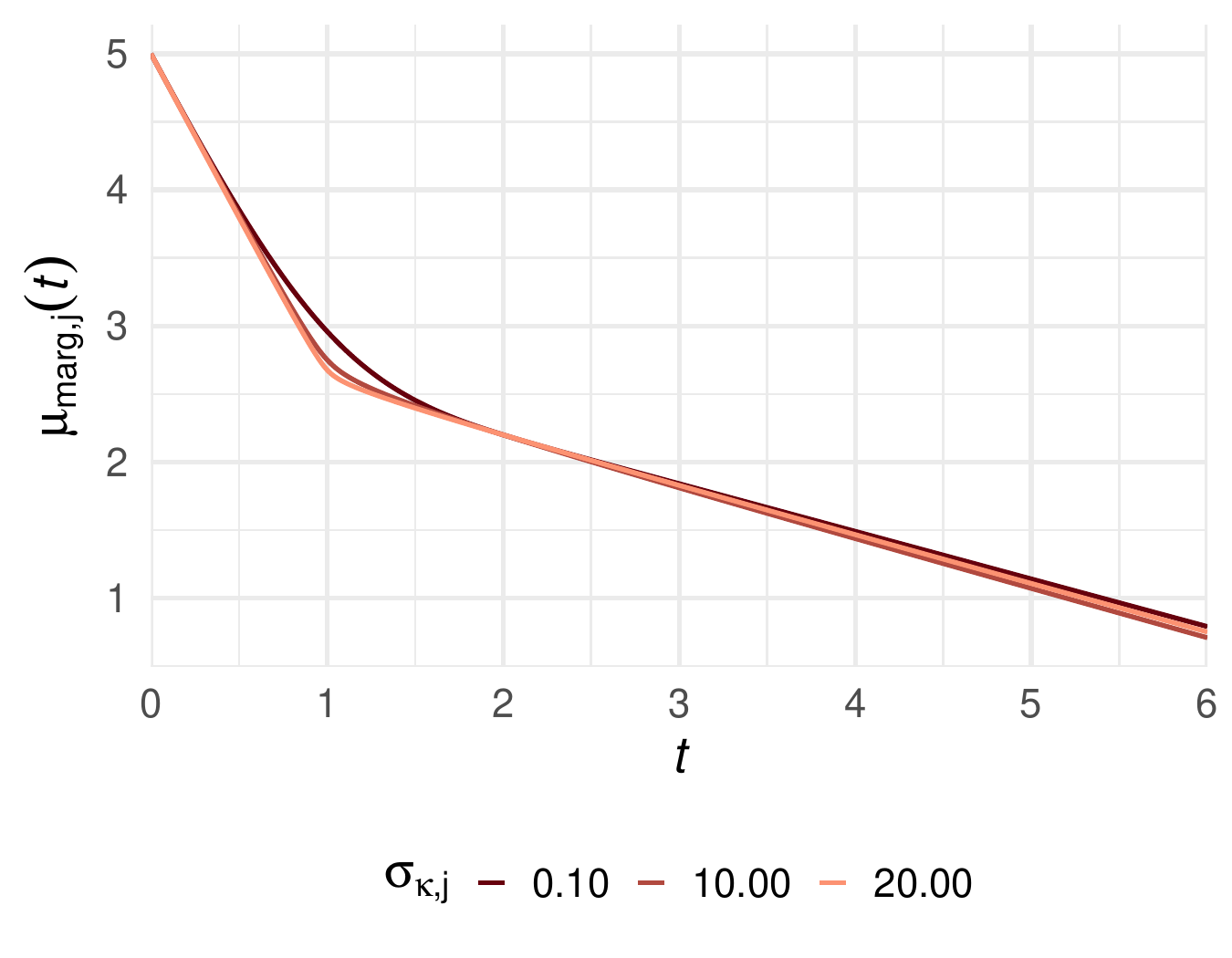}
    
            \vspace{0.6em}
    
            \includegraphics[width=\linewidth]{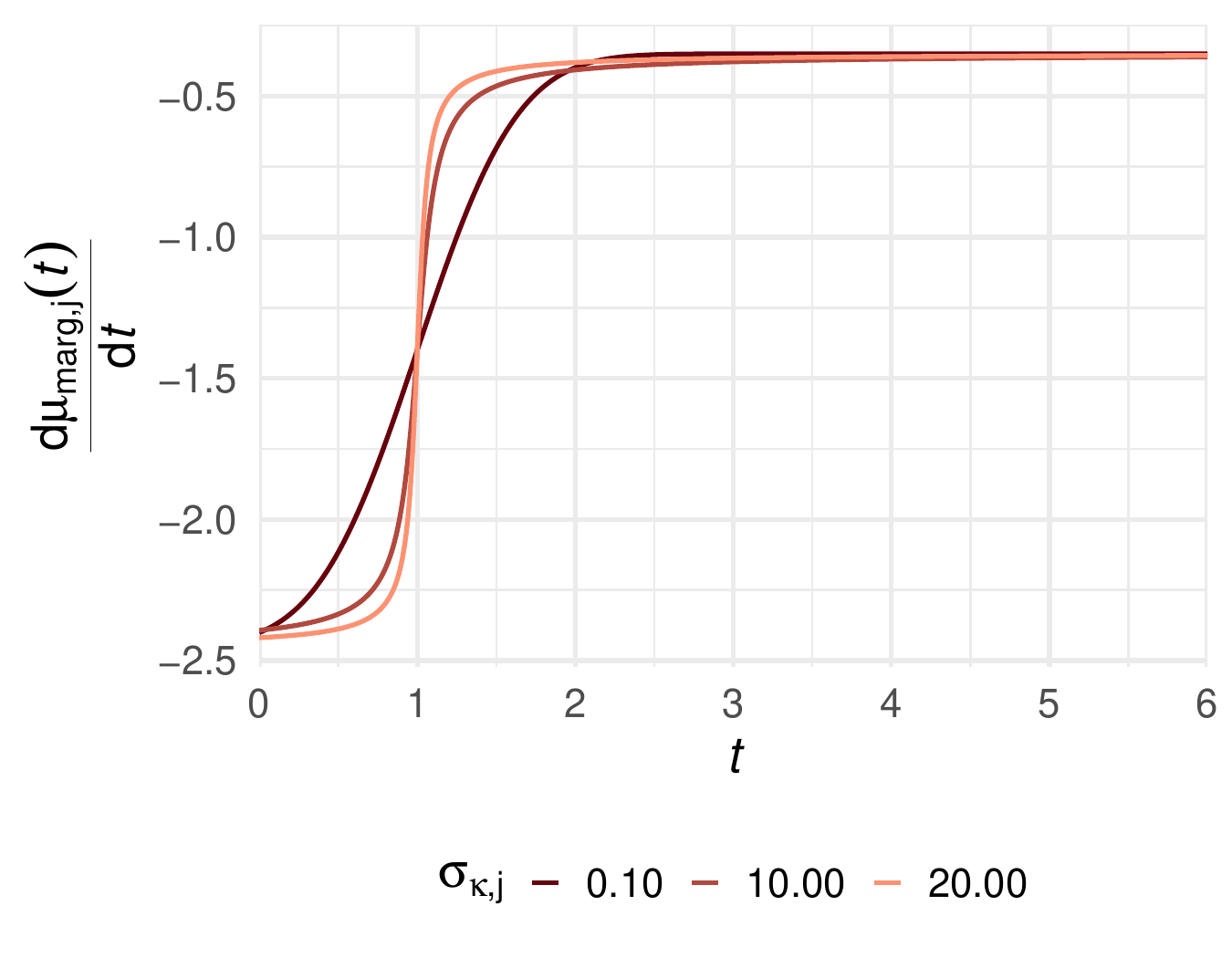}
            \caption{Varying $\sigma_{\kappa}$. Top: mean trajectory. Bottom: instantaneous slope. Fixed $\tau_{ij} \sim \mathrm{TN}\left(1, 0.01^{2}; 0, 6\right)$ with $\sigma_{\kappa,j} \in \left\{0.1, 10, 20\right\}$.}
            \label{fig:slope-sigmak}
        \end{subfigure}
    
        \caption{Illustrative marginal mean trajectories and instantaneous slopes under the smooth changepoint model. In both grouped panels, $\alpha_{0j} = 5$, $\beta_{1j} = -2.45$, $\beta_{2j} = 2.10$, $\mu_{\tau,j} = 1$, and $\mu_{\kappa,j} = 2$. The upper plot in each grouped panel shows $\mu_{\mathrm{marg},j}\left(t\right)$, and the lower plot shows $s_{\mathrm{marg},j}\left(t\right)$.}
        \label{fig:kernel-slopes}
    \end{figure}

    \autoref{fig:kernel-slopes} illustrates how transition timing and transition sharpness enter the marginal summaries. In all panels, we fix $\alpha_{0j} = 5$, $\beta_{1j} = -2.45$, $\beta_{2j} = 2.10$, $\mu_{\tau,j} = 1$, $\mu_{\kappa,j} = 2$, and truncate the changepoint distribution to $\left[0,6\right]$. The upper plot in each grouped panel is the marginal mean trajectory, and the lower plot is the marginal instantaneous slope. In \autoref{fig:slope-sigmat}, the transition sharpness distribution is held fixed while the changepoint dispersion varies. In \autoref{fig:slope-sigmak}, the changepoint distribution is tightly concentrated while the transition sharpness distribution varies. The figure shows that the fitted distributions of $\tau_{ij}$ and $\kappa_{ij}$ can materially alter the population-average transition relative to a plug-in curve; because $\kappa_{ij}$ is an inverse-width parameter, the right-hand panels should be interpreted in terms of sharpness rather than transition width.

    \subsection{Abrupt changepoint model as a limiting case}

    The abrupt two-slope model is obtained by taking $\kappa \to \infty$, so that $h_{\kappa}\left(x\right)$ converges to the hard hinge $x^+$. The corresponding baseline-centered hard-hinge kernel is
    \begin{equation}
        H_j^{\ast}\left(t\right)
        =
        \mathrm{E}_{\tilde{\tau}_j}\left[
        \left(t - \tilde{\tau}_j\right)^+
        -
        \left(-\tilde{\tau}_j\right)^+
        \right].
        \label{eq:H-hard-centered-def}
    \end{equation}
    In the nonnegative changepoint setting used in the ACTG analysis, $\left(-\tilde{\tau}_j\right)^+ = 0$, and this reduces to $\mathrm{E}_{\tilde{\tau}_j}\left[\left(t - \tilde{\tau}_j\right)^+\right]$. With
    \begin{equation}
        z_{t,j}
        =
        \frac{t - \mu_{\tau,j}}{\sigma_{\tau,j}},
        \label{eq:zt-abrupt}
    \end{equation}
    the baseline-centered hard-hinge kernel under the truncated normal law in \eqref{eq:tau-tn-section4}, with $L_{\tau,j}\ge 0$, is
    \begin{equation}
        H_j^{\ast}\left(t\right)
        =
        \begin{cases}
            0,
            &
            t \le L_{\tau,j},\\[8pt]
            \dfrac{
            \left(t - \mu_{\tau,j}\right)
            \left\{\Phi\left(z_{t,j}\right) - \Phi\left(z_{L,j}\right)\right\}
            +
            \sigma_{\tau,j}
            \left\{\phi\left(z_{t,j}\right) - \phi\left(z_{L,j}\right)\right\}
            }{
            Z_{\tau,j}
            },
            &
            L_{\tau,j} < t < R_{\tau,j},\\[12pt]
            t - \mu_{\tau,j}
            +
            \sigma_{\tau,j}
            \dfrac{\phi\left(z_{U,j}\right) - \phi\left(z_{L,j}\right)}{Z_{\tau,j}},
            &
            t \ge R_{\tau,j}.
        \end{cases}
        \label{eq:H-hard-trunc}
    \end{equation}
    Its derivative is the abrupt marginal slope gate
    \begin{equation}
        G_j^{\ast}\left(t\right)
        =
        \frac{\mathrm{d}}{\mathrm{d}t}H_j^{\ast}\left(t\right)
        =
        \begin{cases}
            0,
            &
            t \le L_{\tau,j},\\[8pt]
            \dfrac{\Phi\left(z_{t,j}\right) - \Phi\left(z_{L,j}\right)}{Z_{\tau,j}},
            &
            L_{\tau,j} < t < R_{\tau,j},\\[12pt]
            1,
            &
            t \ge R_{\tau,j}.
        \end{cases}
        \label{eq:G-hard-trunc}
    \end{equation}
    The abrupt marginal mean and slope are therefore
    \begin{equation}
        \mu_{\mathrm{marg},j}^{\ast}\left(t\right)
        =
        \alpha_{0j}
        +
        \beta_{1j}t
        +
        \beta_{2j}
        H_j^{\ast}\left(t\right),
        \label{eq:mu-hard-marg}
    \end{equation}
    and
    \begin{equation}
        s_{\mathrm{marg},j}^{\ast}\left(t\right)
        =
        \beta_{1j}
        +
        \beta_{2j}G_j^{\ast}\left(t\right).
        \label{eq:slope-hard-marg}
    \end{equation}
    If the changepoint distribution is untruncated normal, the corresponding baseline-centered hard-hinge kernel is
    \begin{equation}
    \begin{aligned}
        H_j^{\ast}\left(t\right)
        &=
        \left(t - \mu_{\tau,j}\right)
        \Phi\left(z_{t,j}\right)
        +
        \sigma_{\tau,j}\phi\left(z_{t,j}\right)
        \\
        &\quad
        -
        \left(-\mu_{\tau,j}\right)
        \Phi\left(\frac{-\mu_{\tau,j}}{\sigma_{\tau,j}}\right)
        -
        \sigma_{\tau,j}
        \phi\left(\frac{-\mu_{\tau,j}}{\sigma_{\tau,j}}\right).
    \end{aligned}
        \label{eq:H-hard-limit}
    \end{equation}
    The corresponding slope gate is
    \begin{equation}
        G_j^{\ast}\left(t\right)
        =
        \Phi\left(z_{t,j}\right).
        \label{eq:G-hard-limit}
    \end{equation}
    Thus, even when every subject has a hard changepoint, random changepoint timing can produce a smooth population-average slope.

    \section{Numerical experiment: plug-in vs marginal summaries}
    \label{sec:plugin-numerical}

    We also quantify how much a plug-in representative subject summary can differ from the treatment-arm marginal summary. The purpose is to isolate the effect of heterogeneity in changepoint timing and transition sharpness after the arm-level mean parameters have been fixed.

    We consider a single treatment arm and set the arm-level intercept and slope parameters to the posterior medians from the ACTG~315 smooth changepoint analysis described in \autoref{sec:application}: $\alpha_{0j} = 5.099$, $\beta_{1j} = -4.697$, and $\beta_{2j} = 4.563$. We also fix the pre-truncation locations of the changepoint and transition sharpness distributions to their posterior medians, $\mu_{\tau,j} = 0.436$ and $\mu_{\kappa,j} = 3.608$, respectively. These posterior summaries are reported in \suppautoref{tab:posterior-global}. The pre-truncation locations are not generally equal to the corresponding truncated-normal means $\bar{\tau}_{j}$ and $\bar{\kappa}_{j}$.

    We then vary only the pre-truncation scale parameters, taking
    \begin{equation}
        \sigma_{\tau,j}
        \in
        \left\{0.03, 0.06, 0.093, 0.12, 0.18, 0.25\right\},
        \qquad
        \sigma_{\kappa,j}
        \in
        \left\{0.5, 1.5, 3.377, 5\right\}.
    \end{equation}
    The pair $\sigma_{\tau,j} = 0.093$ and $\sigma_{\kappa,j} = 3.377$ corresponds to the posterior medians of the pre-truncation scale parameters from the ACTG~315 smooth model. Thus, the experiment varies the dispersion of the fitted truncated-normal laws while holding their pre-truncation locations fixed; the implied truncated-normal means may change across scenarios. Throughout the experiment,
    \begin{equation}
        \tilde{\tau}_j
        \sim
        \mathrm{TN}\left(\mu_{\tau,j}, \sigma_{\tau,j}^{2}; 0, 6\right),
        \qquad
        \tilde{\kappa}_j
        \sim
        \mathrm{TN}\left(\mu_{\kappa,j}, \sigma_{\kappa,j}^{2}; 0, \infty\right).
        \label{eq:numerical-experiment-taukappa}
    \end{equation}
    For each $\left(\sigma_{\tau,j},\sigma_{\kappa,j}\right)$ combination, we evaluate the marginal mean trajectory from \eqref{eq:mu-marg-separated} on a dense grid $t_1,\ldots,t_M$ over $\left[0,6\right]$. The inner expectation over $\tilde{\tau}_j$ is computed using the closed-form truncated-normal mean kernel, and the remaining expectation over $\tilde{\kappa}_j$ is computed using the deterministic midpoint quadrature in \eqref{eq:Hbar-quad} with $K = 128$ equal-probability points. We compare this marginal mean to the plug-in mean in \eqref{eq:mu-plugin}, with $\bar{\tau}_j$ and $\bar{\kappa}_j$ recomputed as the corresponding truncated-normal means for each scenario.

    We summarize the discrepancy on the $\log_{10}$ viral load scale using
    \begin{equation}
        \mathrm{RMS}_{\mu}
        =
        \sqrt{
            \frac{1}{M}
            \sum_{m = 1}^{M}
            \left\{
                \mu_{\mathrm{marg},j}\left(t_m\right)
                -
                \mu_{\mathrm{cond},j}\left(t_m\right)
            \right\}^{2}
        },
        \label{eq:rms-mu}
    \end{equation}
    and
    \begin{equation}
        \mathrm{Max}_{\mu}
        =
        \max_{m \in \left\{1,\ldots,M\right\}}
        \left|
            \mu_{\mathrm{marg},j}\left(t_m\right)
            -
            \mu_{\mathrm{cond},j}\left(t_m\right)
        \right|.
        \label{eq:max-mu}
    \end{equation}
    These are discrete $L^2$- and $L^{\infty}$-type summaries of the difference between the two trajectories on the evaluation grid, similar to error criteria used by Ji and M{\"u}ller \citeyearpar{ji2017optimal} for trajectory recovery and by Tan et~al. \citeyearpar{tan2012time} for time-varying effect estimation.

\begin{figure}[!htbp]
        \centering
        \includegraphics[width=\textwidth]{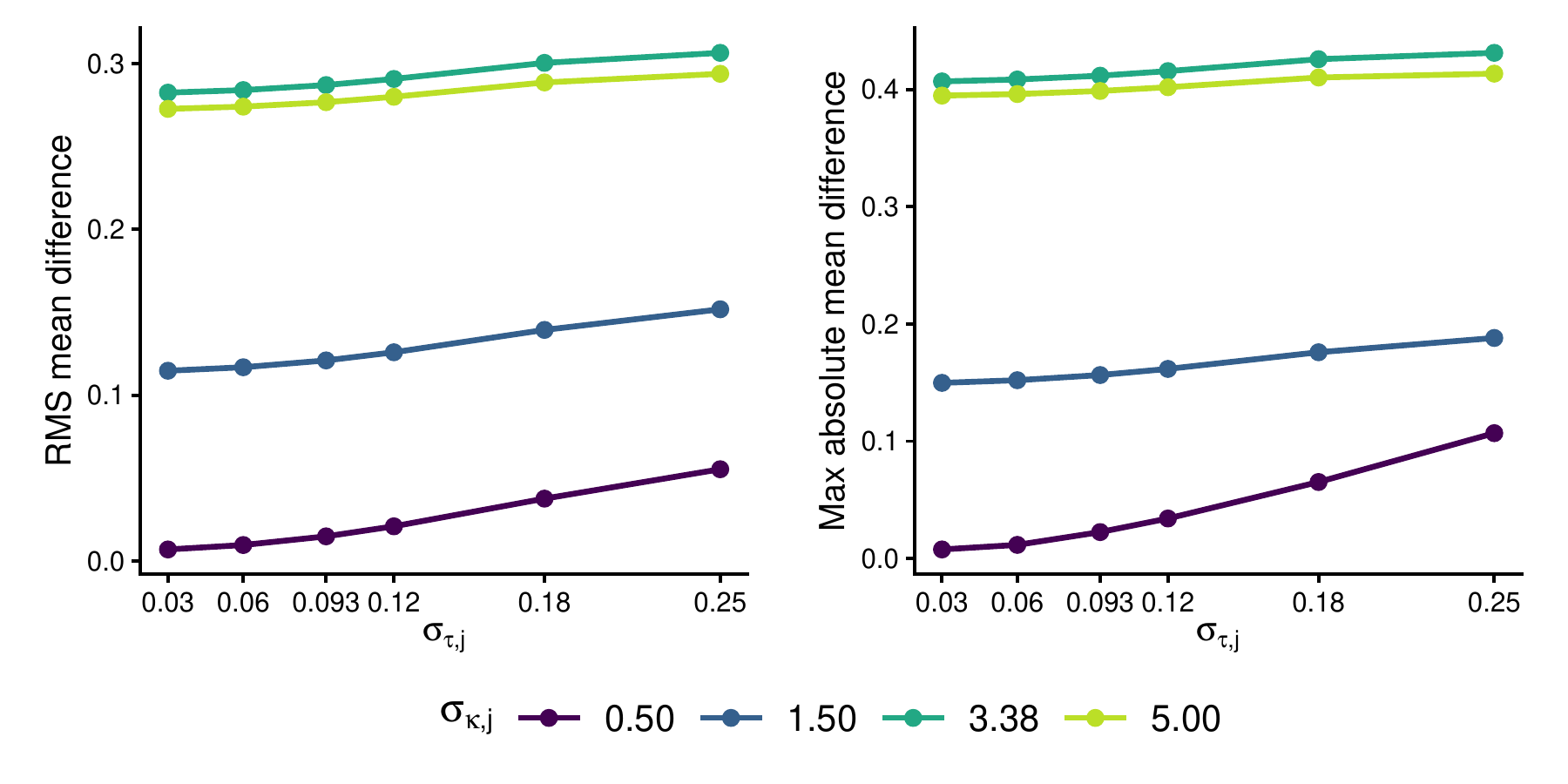}
        \caption{Discrepancy between marginal and plug-in mean trajectories under the smooth changepoint model, as a function of the pre-truncation changepoint scale parameter $\sigma_{\tau,j}$. The left panel shows $\mathrm{RMS}_{\mu}$ and the right panel shows $\mathrm{Max}_{\mu}$, computed over a dense grid of times $t \in \left[0, 6\right]$. Curves correspond to different values of the pre-truncation transition sharpness scale parameter $\sigma_{\kappa,j}$.}
        \label{fig:mean-diff-grid}
    \end{figure}

    \autoref{fig:mean-diff-grid} shows that the discrepancy is negligible when both pre-truncation scale parameters are small, but increases as the changepoint scale parameter increases. The discrepancy also increases substantially as the transition sharpness scale parameter is increased from small values to the ACTG-like value, although it is slightly smaller for the largest value considered. For the ACTG-like setting based on the posterior median pre-truncation scale parameters, $\left(\sigma_{\tau,j} = 0.093, \sigma_{\kappa,j} = 3.377\right)$, the observed discrepancies are $\mathrm{RMS}_{\mu} \approx 0.287$ and $\mathrm{Max}_{\mu} \approx 0.412$ on the $\log_{10}$ scale. A maximum difference of $0.412$ on this scale corresponds to roughly a $2.6$-fold difference in viral load. Thus, the distinction between plug-in and marginal summaries is not just notational: when bend timing and transition sharpness vary across subjects, the plug-in trajectory can materially distort the apparent timing and pace of the population transition.

    \section{ACTG~315 model-based analysis}
    \label{sec:application}

    We now return to the ACTG~315 viral load data described in \autoref{sec:motivating-data}. The outcome is $\log_{10}$ HIV-1 RNA (copies/mL), time is measured in months from baseline, and observations below 100~copies/mL are treated as left-censored at $\log_{10}\left(100\right) = 2$. Because ACTG~315 used a single multidrug antiretroviral regimen, it had a single treatment arm. We retain the arm index $j$ only to keep the notation aligned with the preceding sections.

    \subsection{Model fitting and posterior summaries}

    The primary analysis fits the baseline-anchored smooth random changepoint model in \eqref{eq:subjectcurve}, using the same parameter interpretation as in \autoref{sec:smooth-model}. For comparison, we also fit an abrupt two-slope model in which the centered soft hinge is replaced by the corresponding centered hard hinge and the transition sharpness parameter is omitted.

    Posterior inference for both models is obtained using MCMC in JAGS \citep{PLUMMER2003,PLUMMER2017}. The implementation details are collected in \autoref{app:bayes-implementation}: the skew-$t$ residual distribution is specified in \autoref{app:skewt-resid}, the censoring contribution is given in \autoref{app:censoring}, prior distributions are given in \autoref{app:priors}, and posterior computation is described in \autoref{app:posterior-computation}. Model comparison by Pareto-smoothed importance sampling leave-one-out (PSIS-LOO) is described in \autoref{app:loo}, and posterior predictive residual checks are described in \autoref{app:ppc-residuals}.

    For each retained posterior draw of the smooth model, we compute $\mu_{\mathrm{marg},j}\left(t\right)$ and $s_{\mathrm{marg},j}\left(t\right)$ using \eqref{eq:mu-marg-separated} and \eqref{eq:slope-marg-separated}. These quantities average over the fitted distributions of both $\tau_{ij}$ and $\kappa_{ij}$ and therefore summarize the population-average viral load trajectory and the population-average instantaneous rate of change. We also compute the plug-in trajectory and slope from \eqref{eq:mu-plugin} and \eqref{eq:slope-plugin}. For the abrupt model, we compute the marginal summaries using \eqref{eq:mu-hard-marg} and \eqref{eq:slope-hard-marg}, and the corresponding plug-in summaries by fixing $\tau_{ij}$ at its fitted truncated-normal mean, $\bar{\tau}_j$. All curves are summarized pointwise in time using posterior means and pointwise 95\% credible bands.

    From a reporting perspective, the marginal summaries answer the population question: what is the average underlying viral load trajectory, and how does its instantaneous rate of change evolve over time? The plug-in summaries answer a different conditional question: what trajectory would be obtained for a hypothetical subject whose transition timing and transition sharpness are fixed at typical values?

    \Needspace{6\baselineskip}
    \subsection{Marginal and plug-in trajectories}

    \suppautoref{tab:posterior-global} reports posterior summaries for the abrupt and smooth changepoint models. Under the smooth model, the posterior median early slope is steeply negative, while the estimated slope change is positive, giving a derived late limiting slope close to zero.

\begin{figure}[!htbp]
        \centering
        \includegraphics[width=0.72\textwidth]{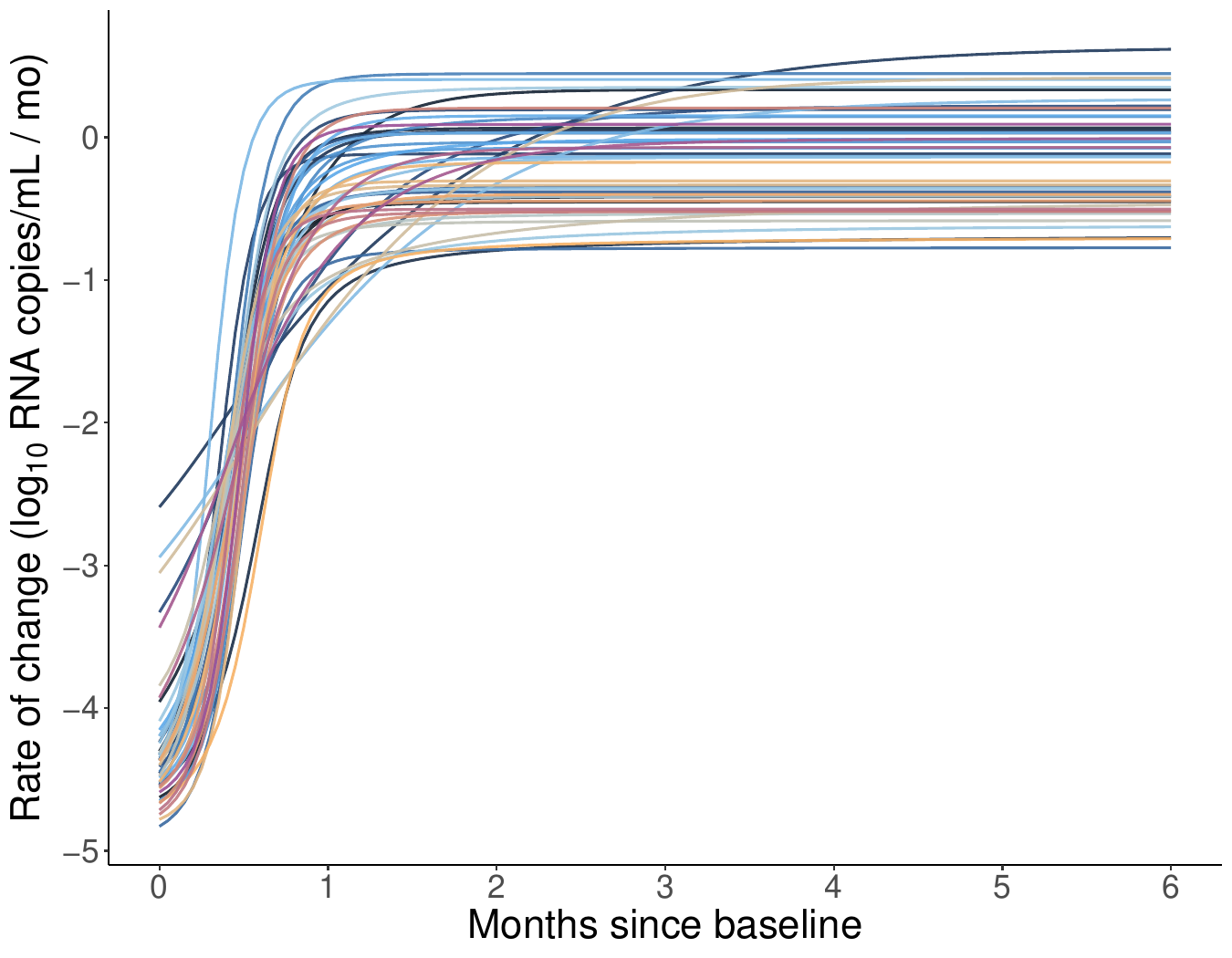}
        \caption{Pointwise posterior mean subject-level instantaneous slopes under the smooth changepoint model. Each curve shows the posterior mean rate of change in $\log_{10}$ HIV-1 RNA (copies/mL per month) over time for one participant.}
        \label{fig:soft-slope-pat}
    \end{figure}

    \autoref{fig:soft-slope-pat} displays posterior mean subject-level instantaneous slope curves under the smooth changepoint model. Most participants show a rapid early decline followed by a substantially flatter long-term phase, but the fitted subject-level transitions differ in timing and apparent sharpness. This subject-level heterogeneity is precisely the feature that the marginal summaries integrate over. As a visual check of the fitted subject-level mean structure, \suppautoref{fig:patient-fits} shows pointwise posterior median fitted latent trajectories under the abrupt and smooth models overlaid on the observed and censored measurements.

\begin{figure}[!htbp]
        \centering
    
        \begin{subfigure}{0.69\textwidth}
            \centering
            \includegraphics[width=\linewidth]{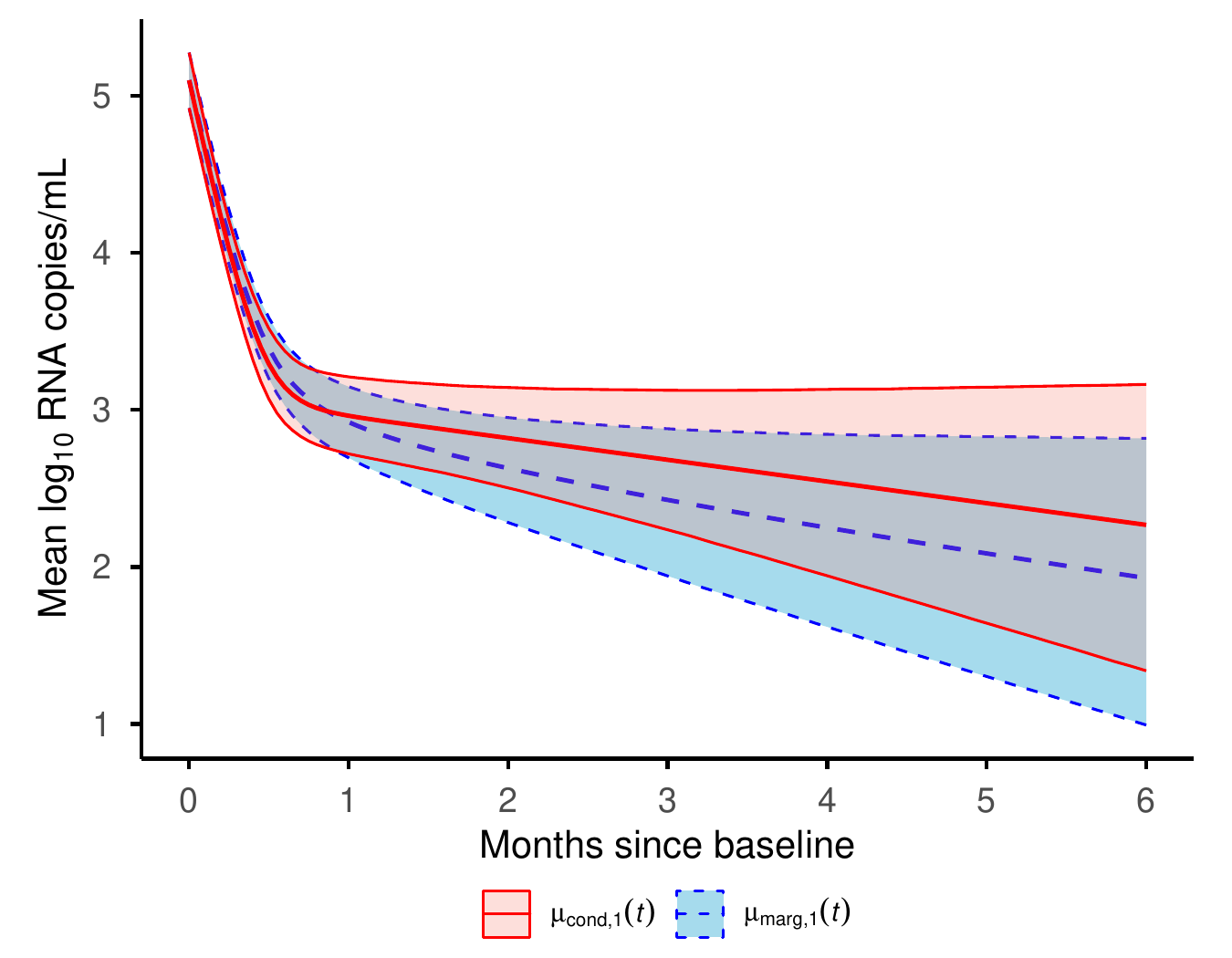}
            \caption{Mean trajectory}
            \label{fig:soft-mean}
        \end{subfigure}
    
        \vspace{1em}
    
        \begin{subfigure}{0.69\textwidth}
            \centering
            \includegraphics[width=\linewidth]{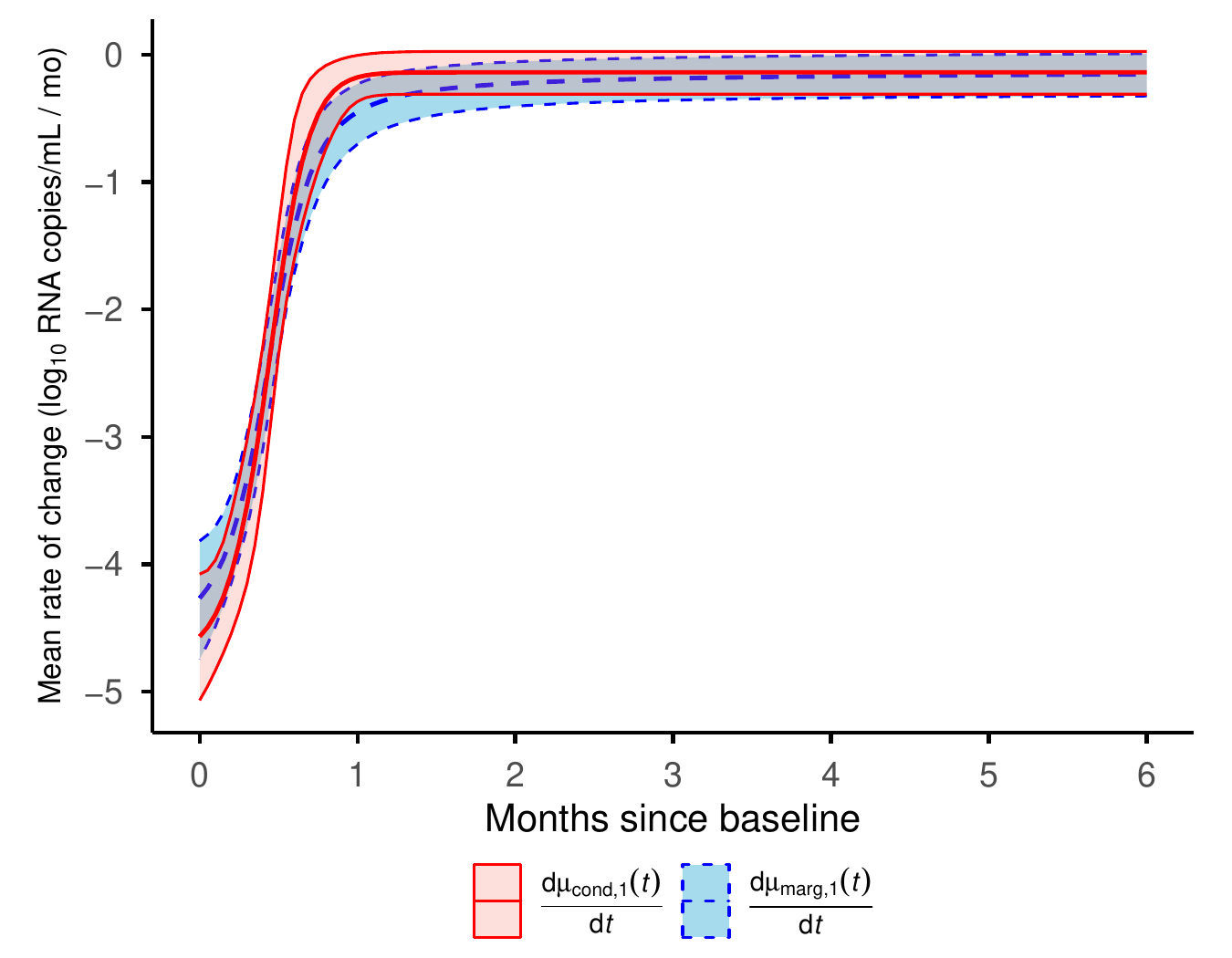}
            \caption{Instantaneous slope}
            \label{fig:soft-slope}
        \end{subfigure}
    
        \caption{Smooth changepoint model. Panel (a) shows mean $\log_{10}$ HIV-1 RNA (copies/mL) over time; panel (b) shows the instantaneous rate of change in $\log_{10}$ HIV-1 RNA (copies/mL per month). In each panel, the dashed curve and band are the marginal summaries, and the solid curve and band are the corresponding conditional plug-in summaries.}
        \label{fig:soft-main}
    \end{figure}

    \suppautoref{fig:bilinear-main} shows the abrupt changepoint summaries, and \autoref{fig:soft-main} shows the corresponding smooth changepoint summaries. The abrupt model already has a smooth marginal slope because changepoint times vary across subjects; therefore, population-level smoothness does not, by itself, imply that individual subjects have smooth transitions. The smooth model adds a subject-level transition sharpness component through $\kappa_{ij}$. In the ACTG~315 analysis, this produces a more gradual population-average transition than the plug-in smooth curve. The plug-in slope moves rapidly from the early steep decline toward the late-phase slope, whereas the marginal slope remains more negative for longer. Consequently, the marginal mean trajectory continues to decline after the plug-in trajectory has already begun to flatten. This is the practical consequence of averaging over subjects whose fitted transitions differ in timing and apparent sharpness.

    \subsection{Predictive model comparison}

    We compared the smooth and abrupt changepoint models using PSIS-LOO cross-validation, summarized by the expected log predictive density and the corresponding leave-one-out (LOO) information criterion, $\mathrm{LOOIC} = -2\widehat{\mathrm{elpd}}_{\mathrm{loo}}$. Because the log-likelihood contributions are evaluated conditional on subject-specific random effects, this observation-level comparison targets the prediction of a held-out measurement conditional on the remaining data from the same subject. It should not be interpreted as a leave-one-subject-out prediction.

    As shown in \autoref{tab:loo-conditional}, the smooth model had a higher estimated expected log predictive density than the abrupt model, with $\widehat{\mathrm{elpd}}_{\mathrm{loo}} = -213.1$ compared with $-234.2$. This corresponds to a lower LOOIC for the smooth model, 426.1 compared with 468.4. The estimated difference in expected log predictive density (smooth minus abrupt) was 21.1 with a standard error of 7.0, indicating better observation-level predictive performance for the smooth changepoint model.

    \begin{longtable}{l@{\hspace{1.4em}}c@{\hspace{0.8em}}c@{\hspace{1.5em}}c@{\hspace{0.8em}}c}
        \caption{LOO cross-validation comparison of observation-level predictive fit. Columns report the approximate expected log predictive density $\widehat{\mathrm{elpd}}_{\mathrm{loo}}$ and the corresponding information criterion $\mathrm{LOOIC} = -2\widehat{\mathrm{elpd}}_{\mathrm{loo}}$, each with its standard error. The estimated difference $\Delta\widehat{\mathrm{elpd}}_{\mathrm{loo}}$ (smooth changepoint minus abrupt changepoint) is 21.1 with standard error 7.0.}\label{tab:loo-conditional}\\
        \hline
        & \multicolumn{2}{c@{\hspace{1.5em}}}{$\boldsymbol{\widehat{\mathrm{elpd}}_{\mathrm{loo}}}$} &
        \multicolumn{2}{c}{$\boldsymbol{\mathrm{LOOIC}}$} \\
        \cline{2-3}\cline{4-5}
        {\bf Changepoint Model} & {\bf Estimate} & {\bf SE} & {\bf Estimate} & {\bf SE} \\
        \hline
        \endfirsthead
        \hline
        & \multicolumn{2}{c@{\hspace{1.5em}}}{$\boldsymbol{\widehat{\mathrm{elpd}}_{\mathrm{loo}}}$} &
        \multicolumn{2}{c}{$\boldsymbol{\mathrm{LOOIC}}$} \\
        \cline{2-3}\cline{4-5}
        {\bf Changepoint Model} & {\bf Estimate} & {\bf SE} & {\bf Estimate} & {\bf SE} \\
        \hline
        \endhead
        \hline
        \endfoot
        \hline
        \endlastfoot
        Abrupt & -234.2 & 22.8 & 468.4 & 45.5 \\*
        Smooth & -213.1 & 21.7 & 426.1 & 43.3 \\
    \end{longtable}

    \subsection{Residual checks}

    \suppautoref{fig:ppc-dharma-qq} and \suppautoref{fig:ppc-pit-qq} show posterior predictive residual diagnostics for the abrupt and smooth changepoint models. These residual diagnostics target calibration under the fitted hierarchical data-generating mechanism, with new subject-specific parameters generated from the fitted random effects and changepoint distributions. By contrast, the PSIS-LOO comparison targets observation-level prediction for an already observed participant, conditional on the subject-specific random effects. The two checks are therefore complementary rather than assessments of the same predictive task.

    The DHARMa QQ plots are broadly consistent with a uniform reference distribution, although some curvature is visible, especially in the middle-to-upper quantiles. The PIT-based normal QQ plots similarly follow the reference line over most of the range, with tail departures. These checks do not show a clear residual calibration advantage for either model. Thus, the empirical support for the smooth model comes primarily from its better observation-level predictive performance, while its substantive advantage lies in its ability to represent a gradual subject-level bend while still producing population-average summaries.

    \section{Discussion}
    \label{sec:discussion}

    This paper distinguishes two summaries that are often conflated in random changepoint modeling. A plug-in trajectory describes a hypothetical representative subject with transition timing and transition sharpness fixed at typical values, whereas a population-average trajectory averages subject-specific curves over the fitted distribution of those quantities. When the timing or shape of the transition varies across subjects, these summaries are generally not the same. The distinction matters because random changepoint timing can smooth the treatment-arm trajectory even when every individual trajectory has an abrupt kink, while subject-level smoothness allows individual trajectories themselves to bend gradually.

    The proposed baseline-anchored normal-convolution hinge accommodates both random changepoint timing and gradual subject-level transitions, while preserving the broken stick interpretation, recovering the abrupt changepoint model as a limiting case, and keeping the population-average mean anchored at baseline. Under the working independence structure, the population-average trajectory and slope can be evaluated through arm-level kernels, with analytic averaging over changepoint timing and a one-dimensional numerical average over transition sharpness.

    The numerical experiment shows that the discrepancy between plug-in and population-average summaries can become substantial when there is heterogeneity in $\tau_{ij}$ or $\kappa_{ij}$. In the ACTG-like setting, the maximum difference between the marginal and plug-in mean trajectories is about $0.412$ on the $\log_{10}$ scale, corresponding to roughly a $2.6$-fold difference in viral load. In the ACTG~315 analysis, the fitted smooth model suggests a population-average slope that remains more negative for longer than the corresponding plug-in slope. These results support reporting population-average summaries when the scientific question concerns the average study population, while reserving plug-in summaries for explicitly conditional representative subject descriptions.
    
    If the scientific target is only the limiting late-phase slope, marginalization over transition timing and transition sharpness is less consequential. The proposed summaries are most relevant when inference concerns the finite-time mean trajectory, the timing and pace of transition, or the instantaneous rate of change during the observed follow-up period.

    Several limitations are important. The deterministic formulas use independence between $\left(\alpha_{0ij},\beta_{1ij},\beta_{2ij}\right)$ and $\left(\tau_{ij},\kappa_{ij}\right)$, as well as independence between $\tau_{ij}$ and $\kappa_{ij}$. More general dependence structures could be accommodated by Monte Carlo averaging under the fitted joint distribution, but would require additional modeling assumptions. The fitted hierarchy is restricted to a single dominant transition and uses parametric truncated normal distributions for changepoint timing and transition sharpness. In applications with multiple distinct phases or subgroup-specific transition patterns, richer transition structures may be needed. In sparse longitudinal data, the timing and width of a transition may be difficult to distinguish fully, so the transition sharpness distribution should be interpreted with caution. Simpler alternatives, such as a common-sharpness model or the abrupt changepoint model, remain useful sensitivity checks when information about transition sharpness heterogeneity is limited.

    Finally, the proposed summaries remain model-based. In ACTG~315, we used a skew-$t$ residual distribution, an explicit likelihood contribution for censoring, and posterior predictive checks. Those checks did not indicate clear systematic residual miscalibration, although some tail departures remained. In other applications, misspecification of the mean structure, random effects distribution, residual family, or censoring mechanism could affect both marginal and plug-in summaries. The main reporting implication is that plug-in curves should be presented as conditional representative subject summaries, not automatically as treatment-arm population summaries.
    
    \BlindedOnly{
    \section*{Acknowledgements}
    
    The authors thank Patrick Schloemer and Andreas Kaiser (Bayer AG) for helpful input on earlier versions of this manuscript.
    
    \section*{Disclosure statement}
    
    No potential conflict of interest was reported by the author(s).
    
    \section*{Funding}
    
    This work was supported by the National Research Foundation of South Africa (Grant number 132383). Opinions expressed and conclusions arrived at are those of the authors and are not necessarily to be attributed to the National Research Foundation.
    }
    
    \AnonymousOnly{
    \section*{Disclosure statement}
    
    No potential conflict of interest was reported by the author(s).
    }
    
    \section*{Data availability statement}
    
    \ifAnonymousManuscript
\else
    The ACTG~315 data used in this study are publicly available as the \texttt{actg315raw} dataset in the \texttt{ushr} R package and from the ACTG~315 longitudinal viral load and CD4+ T-cell data page maintained by Hulin Wu and colleagues at UTHealth. Code for the application, implementation of the proposed method, and the rendered R implementation guide is available at \url{https://github.com/DABURGER1/Marginal-Average-Two-Slope-Modeling}.
    \fi
    
    \section*{Ethical statement}
    
    This study is a secondary analysis of publicly available, de-identified data. No new human participants were recruited for this analysis.

    \appendix

    \section*{Appendix}

    \section{Changepoint-marginalized expectation}
    \label{app:Etau-derivation}
    
    This appendix gives the closed-form calculation used to evaluate the baseline-centered kernel $H_j\left(t \middle| \kappa\right)$ in \eqref{eq:Hj-def-pop}. The derivation is written for a fixed treatment arm $j$ and a fixed transition sharpness value $\kappa > 0$. Let $\tilde{\tau}_j$ denote a generic changepoint from the arm-specific subject-level distribution. For compactness, define the auxiliary raw expectation
    \begin{equation}
        \mathcal{R}_j\left(t \middle| \kappa\right)
        =
        \mathrm{E}_{\tilde{\tau}_j}\left[
        h_{\kappa}\left(t - \tilde{\tau}_j\right)
        \right].
        \label{eq:appendix-R-def}
    \end{equation}
    The baseline-centered kernel used in the main text is then
    \begin{equation}
        H_j\left(t \middle| \kappa\right)
        =
        \mathcal{R}_j\left(t \middle| \kappa\right)
        -
        \mathcal{R}_j\left(0 \middle| \kappa\right).
        \label{eq:appendix-H-from-R}
    \end{equation}
    Let
    \begin{equation}
        \tilde{\tau}_j
        \sim
        \mathrm{TN}\left(\mu_{\tau,j}, \sigma_{\tau,j}^{2}; L_{\tau,j}, R_{\tau,j}\right),
    \end{equation}
    and write $s_j = \sigma_{\tau,j}$. For a given value of $t$ and fixed $\kappa$, define
    \begin{equation}
        m_j
        =
        t - \mu_{\tau,j},
        \qquad
        a_j
        =
        \kappa s_j,
        \qquad
        b_j
        =
        \kappa m_j,
        \qquad
        d_j
        =
        \sqrt{1 + a_j^2}.
    \end{equation}
    This is the same notation as in \eqref{eq:kernel-abd}, with $t$ and $\kappa$ fixed and the arguments suppressed. When the baseline value $\mathcal{R}_j\left(0 \middle| \kappa\right)$ is evaluated, the same definitions are used with $t=0$. With
    \begin{equation}
        z_{L,j}
        =
        \frac{L_{\tau,j} - \mu_{\tau,j}}{s_j},
        \qquad
        z_{U,j}
        =
        \frac{R_{\tau,j} - \mu_{\tau,j}}{s_j},
        \qquad
        Z_{\tau,j}
        =
        \Phi\left(z_{U,j}\right)
        -
        \Phi\left(z_{L,j}\right),
    \end{equation}
    standardizing $x = \left(\tilde{\tau}_j - \mu_{\tau,j}\right)/s_j$ gives $t - \tilde{\tau}_j = m_j - s_j x$. Therefore,
    \begin{equation}
        \mathcal{R}_j\left(t \middle| \kappa\right)
        =
        \frac{1}{Z_{\tau,j}}
        \int_{z_{L,j}}^{z_{U,j}}
        \left\{
        \left(m_j - s_jx\right)\Phi\left(b_j - a_jx\right)
        +
        \frac{1}{\kappa}\phi\left(b_j - a_jx\right)
        \right\}
        \phi\left(x\right)\mathrm{d}x.
        \label{eq:A-target-ij-corrected}
    \end{equation}
    Define the three integrals
    \begin{equation}
        I_{1,j}
        =
        \int_{z_{L,j}}^{z_{U,j}}
        \phi\left(x\right)
        \Phi\left(b_j - a_jx\right)
        \mathrm{d}x,
        \label{eq:I1-app}
    \end{equation}
    \begin{equation}
        I_{2,j}
        =
        \int_{z_{L,j}}^{z_{U,j}}
        x\phi\left(x\right)
        \Phi\left(b_j - a_jx\right)
        \mathrm{d}x,
        \label{eq:I2-app}
    \end{equation}
    and
    \begin{equation}
        I_{3,j}
        =
        \int_{z_{L,j}}^{z_{U,j}}
        \phi\left(x\right)
        \phi\left(b_j - a_jx\right)
        \mathrm{d}x.
        \label{eq:I3-app}
    \end{equation}
    The first integral is obtained from the bivariate normal identity
    \begin{equation}
        \int_{-\infty}^{z}
        \phi\left(x\right)
        \Phi\left(b_j - a_jx\right)
        \mathrm{d}x
        =
        \Phi_2\left(
        z,
        \frac{b_j}{d_j};
        \frac{a_j}{d_j}
        \right),
        \label{eq:bvn-identity-app}
    \end{equation}
    so that
    \begin{equation}
        I_{1,j}
        =
        \Phi_2\left(
        z_{U,j},
        \frac{b_j}{d_j};
        \frac{a_j}{d_j}
        \right)
        -
        \Phi_2\left(
        z_{L,j},
        \frac{b_j}{d_j};
        \frac{a_j}{d_j}
        \right).
        \label{eq:I1-bvn-app}
    \end{equation}
    This immediately gives the slope kernel
    \begin{equation}
        G_j\left(t \middle| \kappa\right)
        =
        \mathrm{E}_{\tilde{\tau}_j}\left[
        \Phi\left(\kappa\left(t - \tilde{\tau}_j\right)\right)
        \right]
        =
        \frac{I_{1,j}}{Z_{\tau,j}},
        \label{eq:G-trunc-app}
    \end{equation}
    which is the expression reported in \eqref{eq:G-trunc-i}.
    
    Integration by parts gives
    \begin{equation}
        I_{2,j}
        =
        \left[
        -\phi\left(z\right)\Phi\left(b_j - a_jz\right)
        \right]_{z=z_{L,j}}^{z_{U,j}}
        -
        a_j I_{3,j}.
        \label{eq:I2-corrected}
    \end{equation}
    Completing the square in the product of the two normal densities gives
    \begin{equation}
        I_{3,j}
        =
        \frac{1}{d_j}
        \phi\left(\frac{b_j}{d_j}\right)
        \left\{
        \Phi\left[
        \left(
        z_{U,j}
        -
        \frac{a_jb_j}{1 + a_j^2}
        \right)d_j
        \right]
        -
        \Phi\left[
        \left(
        z_{L,j}
        -
        \frac{a_jb_j}{1 + a_j^2}
        \right)d_j
        \right]
        \right\}.
        \label{eq:I3-complete-square-app}
    \end{equation}
    Substituting \eqref{eq:I1-bvn-app}, \eqref{eq:I2-corrected}, and \eqref{eq:I3-complete-square-app} into \eqref{eq:A-target-ij-corrected} yields
    \begin{equation}
        \mathcal{R}_j\left(t \middle| \kappa\right)
        =
        \frac{1}{Z_{\tau,j}}
        \left\{
        m_j I_{1,j}
        +
        s_j
        \left[
        \phi\left(z_{U,j}\right)\Phi\left(b_j - a_jz_{U,j}\right)
        -
        \phi\left(z_{L,j}\right)\Phi\left(b_j - a_jz_{L,j}\right)
        \right]
        +
        \left(\frac{1}{\kappa} + \kappa s_j^2\right)I_{3,j}
        \right\}.
        \label{eq:Etau-hkappa-main-i}
    \end{equation}
    Combining \eqref{eq:Etau-hkappa-main-i} with \eqref{eq:appendix-H-from-R} gives the closed-form baseline-centered kernel $H_j\left(t \middle| \kappa\right)$ used in the main text.
    
    In the untruncated case, $z_{L,j}=-\infty$, $z_{U,j}=\infty$, and $Z_{\tau,j}=1$. The boundary term in \eqref{eq:Etau-hkappa-main-i} vanishes, $I_{1,j}=\Phi\left(b_j/d_j\right)$, and $I_{3,j}=d_j^{-1}\phi\left(b_j/d_j\right)$. Hence
    \begin{equation}
        \mathcal{R}_j\left(t \middle| \kappa\right)
        =
        m_j
        \Phi\left(\frac{b_j}{d_j}\right)
        +
        \left(\frac{1}{\kappa} + \kappa s_j^2\right)
        \frac{1}{d_j}
        \phi\left(\frac{b_j}{d_j}\right),
        \label{eq:Etau-untrunc-app}
    \end{equation}
    where
    \begin{equation}
        \frac{b_j}{d_j}
        =
        \frac{m_j}{\sqrt{s_j^2 + \kappa^{-2}}},
        \qquad
        \left(\frac{1}{\kappa} + \kappa s_j^2\right)
        \frac{1}{d_j}
        =
        \sqrt{s_j^2 + \kappa^{-2}}.
    \end{equation}
    Centering \eqref{eq:Etau-untrunc-app} at $t=0$ gives \eqref{eq:Etau-untrunc-i}. Differentiating gives
    \begin{equation}
        G_j\left(t \middle| \kappa\right)
        =
        \Phi\left(
        \frac{m_j}{\sqrt{s_j^2 + \kappa^{-2}}}
        \right),
        \label{eq:G-untrunc-app}
    \end{equation}
    which is \eqref{eq:G-untrunc-i}.

    \section{Bayesian implementation for ACTG~315}
    \label{app:bayes-implementation}

    This appendix gives the likelihood, censoring mechanism, residual distribution, priors, and posterior computation used in \autoref{sec:application}. For ACTG~315, there is a single treatment arm (we retain the arm index $j$ for consistency with \autoref{sec:smooth-model} and \autoref{sec:marginal-trajectories}). The subject-level mean structure, the smooth hinge function $h_{\kappa}\left(\cdot\right)$, the interpretation of $\left(\alpha_{0ij},\beta_{1ij},\beta_{2ij}\right)$, and the distributions for $\tau_{ij}$ and $\kappa_{ij}$ are as defined in \autoref{sec:smooth-model}. In the abrupt model, the centered soft-hinge contribution in \eqref{eq:subjectcurve} is replaced by the corresponding centered hard-hinge contribution, which reduces to $\left(t - \tau_{ij}\right)^{+}$ because $\tau_{ij}\ge 0$, and the transition sharpness parameter $\kappa_{ij}$ is omitted.

    \subsection{Skew-\textbf{\textit{t}} residual distribution}
    \label{app:skewt-resid}

    We instantiate the generic residual model in \eqref{eq:resid} for the ACTG~315 analysis by taking $f_{\varepsilon}$ to be the skew-$t$ density proposed by Hansen \citeyearpar{hansen1994autoregressive}, with arm-specific residual parameters $\boldsymbol{\theta}_j = \left(\sigma_j,\Lambda_j,\nu_j\right)$.

    Let $Y_{ijk}$ denote the observed $\log_{10}$ viral load for subject $i$ in arm $j$ at time $t_{ijk}$ and let $Y_{ijk}^\ast$ denote the latent uncensored value. Conditional on the subject-specific parameters and on the arm-level residual parameters, we model
    \begin{equation}
        \left.
        Y_{ijk}^\ast
        \middle|
        \alpha_{0ij}, \beta_{1ij}, \beta_{2ij}, \tau_{ij}, \kappa_{ij},
        \sigma_j, \Lambda_j, \nu_j
        \right.
        \sim
        \mathrm{Skew}\text{-}t
        \left(
        \mu_{ijk},
        \sigma_j,
        \Lambda_j,
        \nu_j
        \right),
        \label{eq:skewt-app-appendix}
    \end{equation}
    where $\mu_{ijk} = \mu_{ij}\left(t_{ijk}\right)$ with $\mu_{ij}\left(t\right)$ given in \eqref{eq:subjectcurve} for the smooth model and in the corresponding hard-hinge form for the abrupt model, $\sigma_j > 0$ is a residual scale parameter for arm $j$, $\Lambda_j \in \left(-1,1\right)$ controls asymmetry, and $\nu_j > 2$ controls tail thickness.

    Write the standardized residual
    \begin{equation}
        z_{ijk}
        =
        \frac{Y_{ijk}^\ast - \mu_{ijk}}{\sigma_j}.
        \end{equation}
        Following the piecewise skew-$t$ parameterization used in the analysis, define
        \begin{equation}
        C_{c,j}
        =
        \frac{\Gamma\left(\frac{\nu_j + 1}{2}\right)}
        {\sqrt{\pi \left(\nu_j - 2\right)}
         \Gamma\left(\frac{\nu_j}{2}\right)},
        \qquad
        A_j
        =
        \frac{4 \Lambda_j C_{c,j} \left(\nu_j - 2\right)}{\nu_j - 1},
        \qquad
        B_j
        =
        \sqrt{1 + 3\Lambda_j^{2} - A_j^{2}},
    \end{equation}
    and
    \begin{equation}
        K_{\mathrm{knot},j}
        =
        -\frac{A_j}{B_j},
        \qquad
        K_{\mathrm{fac},j}
        =
        \sqrt{\frac{\nu_j}{\nu_j - 2}}.
    \end{equation}
    Let
    \begin{equation}
        \tilde{z}_{ijk}
        =
        B_j z_{ijk} + A_j.
    \end{equation}
    Define
    \begin{equation}
        g_{L,j}\left(z_{ijk}\right)
        =
        1
        +
        \frac{1}{\nu_j - 2}
        \left(
        \frac{\tilde{z}_{ijk}}{1 - \Lambda_j}
        \right)^{2},
        \qquad
        g_{R,j}\left(z_{ijk}\right)
        =
        1
        +
        \frac{1}{\nu_j - 2}
        \left(
        \frac{\tilde{z}_{ijk}}{1 + \Lambda_j}
        \right)^{2}.
    \end{equation}
    The resulting skew-$t$ probability density function is
    \begin{equation}
        f_{\mathrm{ST},j}\left(
        Y_{ijk}^\ast \left. \right|
        \mu_{ijk}, \sigma_j, \Lambda_j, \nu_j
        \right)
        =
        \frac{B_j C_{c,j}}{\sigma_j}
        \times
        \begin{cases}
        g_{L,j}\left(z_{ijk}\right)^{-\left(\nu_j + 1\right)/2},
        & z_{ijk} \le K_{\mathrm{knot},j}, \\[6pt]
        g_{R,j}\left(z_{ijk}\right)^{-\left(\nu_j + 1\right)/2},
        & z_{ijk} > K_{\mathrm{knot},j}.
        \end{cases}
        \label{eq:skewt-pdf-app-appendix}
    \end{equation}
    For any threshold $u$, we define
    \begin{equation}
        z_{ijk}^{\left(u\right)}
        =
        \frac{u - \mu_{ijk}}{\sigma_j},
    \end{equation}
    and write $T_{\nu_j}\left(\cdot\right)$ for the CDF of a standard Student $t$ distribution with $\nu_j$ degrees of freedom. The skew-$t$ CDF is
    \begin{equation}
        F_{\mathrm{ST},j}\left(
        u \left. \right|
        \mu_{ijk}, \sigma_j, \Lambda_j, \nu_j
        \right)
        =
        \begin{cases}
            \left(1 - \Lambda_j\right)
            T_{\nu_j}\left(
            K_{\mathrm{fac},j}
            \dfrac{B_j z_{ijk}^{\left(u\right)} + A_j}{1 - \Lambda_j}
            \right),
            &
            z_{ijk}^{\left(u\right)} \le K_{\mathrm{knot},j},
            \\[10pt]
            1
            -
            \left(1 + \Lambda_j\right)
            T_{\nu_j}\left(
            K_{\mathrm{fac},j}
            \dfrac{-\left(B_j z_{ijk}^{\left(u\right)} + A_j\right)}{1 + \Lambda_j}
            \right),
            &
            z_{ijk}^{\left(u\right)} > K_{\mathrm{knot},j}.
        \end{cases}
        \label{eq:skewt-cdf-app-appendix}
    \end{equation}
    This residual specification allows arm-specific heavy tails (small $\nu_j$), arm-specific asymmetry ($\Lambda_j \ne 0$), and different residual scales across arms, and reduces to a symmetric standardized Student-$t$ distribution for $\Lambda_j = 0$.

    \subsection{Censoring at the lower limit of quantification}
    \label{app:censoring}

    Let $u_{ijk}$ denote the LLOQ on the $\log_{10}$ scale (for ACTG~315, $u_{ijk} = 2$ for all $i,j,k$). We observe
    \begin{equation}
        Y_{ijk}
        =
        \max\left(Y_{ijk}^\ast, u_{ijk}\right),
    \end{equation}
    and define $c_{ijk} = 1$ if $Y_{ijk}^\ast$ is observed (uncensored) and $c_{ijk} = 0$ if the measurement is reported only as below quantification (left censored at $u_{ijk}$). The likelihood contribution of observation $\left(i,j,k\right)$ is
    \begin{equation}
        \ell_{ijk}
        =
        \left[
        f_{\mathrm{ST},j}\left(
        Y_{ijk} \left. \right|
        \mu_{ijk}, \sigma_j, \Lambda_j, \nu_j
        \right)
        \right]^{c_{ijk}}
        \left[
        F_{\mathrm{ST},j}\left(
        u_{ijk} \left. \right|
        \mu_{ijk}, \sigma_j, \Lambda_j, \nu_j
        \right)
        \right]^{1 - c_{ijk}}.
        \label{eq:likelihood-ijk-appendix}
    \end{equation}
    The subject-level likelihood is
    \begin{equation}
        \mathcal{L}_{ij}
        =
        \prod_{k=1}^{n_{ij}}
        \ell_{ijk},
        \label{eq:likelihood-subject-appendix}
    \end{equation}
    and the conditional data likelihood has the same product form as in \eqref{eq:lik-full}, with the observation-level contribution replaced by the censored skew-$t$ contribution in \eqref{eq:likelihood-ijk-appendix}.
    
    \subsection{Priors on residual, changepoint, and smoothness terms}
    \label{app:priors}

    Arm-specific residual parameters are given weakly informative priors:
    \begin{equation}
        \label{eq:priors-resid-appendix}
        \begin{aligned}
            \sigma_j      &\sim \mathrm{t}_3^{+}\left(0, s_{\sigma}\right), \\
            \Lambda_j     &\sim \mathrm{Uniform}\left(-1,1\right), \\
            \nu_j         &= 2 + \xi_j, \\
            \xi_j         &\sim \mathrm{Gamma}\left(2, \eta_j\right), \\
            \eta_j        &\sim \mathrm{Exp}\left(1\right).
        \end{aligned}
    \end{equation}
    Here, $\mathrm{t}_3^{+}\left(0, s_{\sigma}\right)$ denotes a Student-$t$ distribution with $3$ degrees of freedom, location $0$, and scale $s_{\sigma}$, truncated to $\left(0,\infty\right)$ (a half-$t_3$ prior). In the ACTG~315 implementation, we take $s_{\sigma} = \sqrt{2}$. The notation $\mathrm{Gamma}\left(2, \eta_j\right)$ refers to a gamma distribution with shape $2$ and rate parameter $\eta_j$, and $\mathrm{Exp}\left(1\right)$ denotes an exponential distribution with rate $1$. Integrating out $\xi_j$ and $\eta_j$ yields the implied marginal prior for $\nu_j$, with density
    \begin{equation}
        f_{\nu_j}\left(\nu\right)
        =
        \frac{2\left(\nu - 2\right)}{\left(\nu - 1\right)^{3}},
        \qquad
        \nu > 2,
    \end{equation}
    which provides a heavy-tailed prior on the degrees of freedom, in line with the recommendations of Ju{\'a}rez and Steel \citeyearpar{JUAREZ2010}.

    For the changepoint and transition sharpness parameters, we use the same truncated normal structure introduced in \autoref{sec:smooth-model} and assign weakly informative hyperpriors to the arm-specific location and scale parameters. For each arm $j$,
    \begin{equation}
        \mu_{\tau,j} \sim \mathcal{N}\left(0, 1000\right),
        \quad
        \sigma_{\tau,j} \sim \mathrm{t}_3^{+}\left(0, s_{\sigma}\right),
        \quad
        \mu_{\kappa,j} \sim \mathcal{N}\left(0, 1000\right),
        \quad
        \sigma_{\kappa,j} \sim \mathrm{t}_3^{+}\left(0, s_{\sigma}\right),
        \label{eq:priors-taukappa-appendix}
    \end{equation}
    so that the same half-$t_3$ scale prior is reused for the changepoint and smoothness standard deviations. The individual changepoints and transition sharpness parameters satisfy
    \begin{equation}
        \tau_{ij}
        \sim
        \mathrm{TN}\left(\mu_{\tau,j}, \sigma_{\tau,j}^{2}; 0, \tau_{\max}\right),
        \qquad
        \kappa_{ij}
        \sim
        \mathrm{TN}\left(\mu_{\kappa,j}, \sigma_{\kappa,j}^{2}; 0, \infty\right),
    \end{equation}
    independently across subjects, and independently of each other within subject, conditionally on $\mu_{\tau,j}$, $\sigma_{\tau,j}^{2}$, $\mu_{\kappa,j}$, and $\sigma_{\kappa,j}^{2}$. In the ACTG~315 analysis, $\tau_{\max}=6$ months. In the abrupt changepoint model, $\kappa_{ij}$ and its hyperparameters $\left(\mu_{\kappa,j},\sigma_{\kappa,j}^{2}\right)$ are omitted.

    For the random effects distribution of $\left(\alpha_{0ij},\beta_{1ij},\beta_{2ij}\right)$ in \eqref{eq:re}, we place weakly informative priors on the arm-level means and covariance. Specifically,
    \begin{equation}
        \alpha_{0j},\ \beta_{1j},\ \beta_{2j}
        \sim
        \mathcal{N}\left(0, 1000\right)
    \end{equation}
    independently, and we parameterize the precision matrix $\boldsymbol{\Sigma}_{v,j}^{-1}$ via its $3 \times 3$ lower triangular Cholesky factor $\boldsymbol{L}_j$, so that $\boldsymbol{\Sigma}_{v,j}^{-1} = \boldsymbol{L}_j \boldsymbol{L}_j^{\top}$ with
    \begin{equation}
        \boldsymbol{L}_j
        =
        \begin{pmatrix}
            L_{j,11} & 0 & 0 \\
            L_{j,21} & L_{j,22} & 0 \\
            L_{j,31} & L_{j,32} & L_{j,33}
        \end{pmatrix}.
    \end{equation}
    The diagonal entries $L_{j,11},L_{j,22},L_{j,33}$ receive independent $\mathrm{t}_3^{+}\left(0, s_{\sigma}\right)$ priors and the strict subdiagonal entries $L_{j,21},L_{j,31},L_{j,32}$ receive independent $\mathcal{N}\left(0,1000\right)$ priors. Conditional on $\left(\alpha_{0j},\beta_{1j},\beta_{2j},\boldsymbol{\Sigma}_{v,j}\right)$, the random effects $\left(\alpha_{0ij},\beta_{1ij},\beta_{2ij}\right)$ are independent across subjects.

    \subsection{Posterior and computation}
    \label{app:posterior-computation}

    Let $\boldsymbol{\Theta}$ collect the fixed effects, random effects, and hyperparameters across all arms and subjects. For each arm $j$, this includes the arm-level means for $\left(\alpha_{0ij},\beta_{1ij},\beta_{2ij}\right)$, the lower-triangular Cholesky factor $\boldsymbol{L}_j$ of the random effects precision matrix, the parameters of the truncated normal laws for $\tau_{ij}$ and (in the smooth model) $\kappa_{ij}$, and the residual parameters $\sigma_j,\Lambda_j,\nu_j$. Writing
    \begin{equation}
        \boldsymbol{Y}
        =
        \left\{Y_{ijk}\right\}_{i,j,k},
        \qquad
        \boldsymbol{c}
        =
        \left\{c_{ijk}\right\}_{i,j,k},
        \qquad
        \boldsymbol{t}
        =
        \left\{t_{ijk}\right\}_{i,j,k},
        \qquad
        \boldsymbol{u}
        =
        \left\{u_{ijk}\right\}_{i,j,k},
    \end{equation}
    and defining
    \begin{equation}
    \begin{aligned}
        \boldsymbol{\theta}_j
        &=
        \left(\sigma_j,\Lambda_j,\nu_j\right)^{\top},\\
        \boldsymbol{v}_{ij}
        &=
        \left(\alpha_{0ij},\beta_{1ij},\beta_{2ij}\right)^{\top},
        \qquad
        \boldsymbol{v}_{j}
        =
        \left(\alpha_{0j},\beta_{1j},\beta_{2j}\right)^{\top},
    \end{aligned}
    \end{equation}
    the joint posterior density for the smooth changepoint model is
    \begin{equation}
        \label{eq:posterior-appendix}
        \begin{aligned}
            p\left(
              \boldsymbol{\Theta}
              \middle|
              \boldsymbol{Y},\boldsymbol{c},\boldsymbol{t},\boldsymbol{u}
            \right)
            &\propto
            \prod_{j=1}^{J}
            \prod_{i=1}^{N_j}
            \prod_{k=1}^{n_{ij}}
            \Bigg\{
                f_{\mathrm{ST},j}
                \left(
                    Y_{ijk}
                    \middle|
                    \mu_{ijk},
                    \boldsymbol{\theta}_j
                \right)^{c_{ijk}}
                \\
                &\hspace{7em}\times
                F_{\mathrm{ST},j}
                \left(
                    u_{ijk}
                    \middle|
                    \mu_{ijk},
                    \boldsymbol{\theta}_j
                \right)^{1 - c_{ijk}}
            \Bigg\}
            \\
            &\quad\times
            \prod_{j=1}^{J}
            \Bigg[
                p\left(\boldsymbol{v}_{j}\right)
                p\left(\boldsymbol{L}_j\right)
                p\left(\mu_{\tau,j},\sigma_{\tau,j}\right)
                p\left(\mu_{\kappa,j},\sigma_{\kappa,j}\right)
                p\left(\boldsymbol{\theta}_j\right)
                \\
                &\hspace{5em}\times
                \prod_{i=1}^{N_j}
                p\left(
                    \boldsymbol{v}_{ij}
                    \middle|
                    \boldsymbol{v}_{j},\boldsymbol{L}_j
                \right)
                p\left(
                    \tau_{ij}
                    \middle|
                    \mu_{\tau,j},\sigma_{\tau,j}^{2}
                \right)
                \\
                &\hspace{8em}\times
                p\left(
                    \kappa_{ij}
                    \middle|
                    \mu_{\kappa,j},\sigma_{\kappa,j}^{2}
                \right)
            \Bigg],
        \end{aligned}
    \end{equation}
    where $\mu_{ijk} = \mu_{ij}\left(t_{ijk}\right)$ is given by \eqref{eq:subjectcurve} in the smooth model, and
    $f_{\mathrm{ST},j}\left(\cdot \middle| \mu_{ijk},\boldsymbol{\theta}_j\right)$ and
    $F_{\mathrm{ST},j}\left(\cdot \middle| \mu_{ijk},\boldsymbol{\theta}_j\right)$ denote the skew-$t$ density and CDF defined in \eqref{eq:skewt-pdf-app-appendix} and \eqref{eq:skewt-cdf-app-appendix}. The matrix $\boldsymbol{L}_j$ is the lower-triangular Cholesky factor of the random effects precision matrix, so that $\boldsymbol{\Sigma}_{v,j}^{-1} = \boldsymbol{L}_j\boldsymbol{L}_j^{\top}$. The conditional density $p\left(\boldsymbol{v}_{ij} \middle| \boldsymbol{v}_{j},\boldsymbol{L}_j\right)$ is the multivariate normal random effects density with mean $\boldsymbol{v}_{j}$ and covariance matrix $\boldsymbol{\Sigma}_{v,j}$ implied by $\boldsymbol{L}_j$. The prior factors follow \eqref{eq:priors-resid-appendix} and \eqref{eq:priors-taukappa-appendix}, together with the random effects specification in \eqref{eq:re}; the conditional densities for $\tau_{ij}$ and $\kappa_{ij}$ are the corresponding truncated normal densities. In the abrupt changepoint model, the hard-hinge counterpart of \eqref{eq:subjectcurve} is used and the terms involving $\kappa_{ij}$ and $\left(\mu_{\kappa,j},\sigma_{\kappa,j}^{2}\right)$ are omitted. For $\nu_j$, this hierarchical construction is equivalent to using the implied marginal prior with density $f_{\nu_j}\left(\nu\right)$ after integrating out the auxiliary rate parameter $\eta_j$ and using $\xi_j=\nu_j-2$.

    Posterior inference is obtained by MCMC in JAGS \citep{PLUMMER2003,PLUMMER2017}. For each fitted model, we ran 10 parallel chains with 1000 adaptive iterations and 25000 burn-in iterations per chain. The sampling phase used a thinning interval of 100 and retained 4000 post-burn-in draws per chain, giving 40000 retained posterior draws for each model. Computations were performed on a laptop with a 13th-generation Intel Core i7-1365U processor, 10 physical cores, 12 logical processors, and 32~GB RAM. On this computing environment, the abrupt and smooth changepoint models required approximately 9.1 and 10.1 hours of wall-clock runtime, respectively. Initial values for the subject-specific parameters are obtained from simple subject-level linear fits. For each participant, we fit separate ordinary least squares regressions of censored $\log_{10}$ HIV-1 RNA on time measured in months before and after 1 month, using uncensored observations when available. The intercept from the early-time fit initializes $\alpha_{0ij}$, the corresponding early slope initializes $\beta_{1ij}$, and the difference between the late and early slopes initializes $\beta_{2ij}$. We set $\tau_{ij} = 1$ month for all subjects, initialize $\kappa_{ij}$ at moderate positive values in the smooth model, and initialize the arm-level means $\left(\alpha_{0j}, \beta_{1j}, \beta_{2j}\right)$ at the sample averages of these subject-level estimates.

    Convergence and mixing are assessed using standard diagnostics, including trace plots and the potential scale reduction factor of Gelman and Rubin \citeyearpar{gelman1992inference}, and posterior summaries are based on all retained post-burn-in draws from all chains. For the ACTG~315 application, the deterministic average over transition sharpness used $K = 32$ equal-probability quadrature points for each retained posterior draw. From each retained posterior draw of the smooth model, we compute the marginal mean trajectory $\mu_{\mathrm{marg},j}\left(t\right)$ and marginal slope $s_{\mathrm{marg},j}\left(t\right)$ defined in \autoref{sec:marginal-trajectories}. These functions average over the fitted distributions of changepoint times and transition sharpness parameters and therefore describe, respectively, the population-average viral load path and the population-average instantaneous rate of change at time $t$ in the ACTG~315 study population. We summarize these functions pointwise in $t$ using posterior means and 95\% credible bands, and we compare them with the conditional plug-in summaries defined in \eqref{eq:mu-plugin} and \eqref{eq:slope-plugin} to clarify the distinction between population-average and representative subject interpretations.

    \subsection{Leave-one-out cross-validation for model comparison}
    \label{app:loo}

    To compare the smooth changepoint model with the abrupt two-slope model, we used PSIS-LOO cross-validation based on the observation-level log-likelihood under each fitted model, implemented via the \texttt{loo} package \citep{vehtari2024LOO}.

    Let $Y_{ijk}$ denote the $k$\textsuperscript{th} observed viral load value on the $\log_{10}$ scale for subject $i$ in arm $j$, with corresponding quantification limit $u_{ijk}$ and censoring indicator $c_{ijk}$, where $c_{ijk}=1$ for uncensored measurements and $c_{ijk}=0$ for values reported only as below the LLOQ. For a given model and posterior draw $s=1,\ldots,S$, let $\mu_{ijk}^{\left(s\right)}$ be the implied mean trajectory at the observation time for $Y_{ijk}$, conditional on the corresponding sampled subject-specific parameters, and let $\boldsymbol{\theta}_{j}^{\left(s\right)}$ collect the residual parameters $\left(\sigma_{j},\Lambda_{j},\nu_{j}\right)$ for arm $j$. Using the skew-$t$ density $f_{\mathrm{ST},j}$ and CDF $F_{\mathrm{ST},j}$ defined in \autoref{app:skewt-resid}, the observation-level log-likelihood contribution is
    \begin{equation}
      \ell_{ijk}^{\left(s\right)}
      =
      c_{ijk}
      \log
      \left[
        f_{\mathrm{ST},j}
        \left(
          Y_{ijk}
          \middle|
          \mu_{ijk}^{\left(s\right)},
          \boldsymbol{\theta}_{j}^{\left(s\right)}
        \right)
      \right]
      +
      \left(1-c_{ijk}\right)
      \log
      \left[
        F_{\mathrm{ST},j}
        \left(
          u_{ijk}
          \middle|
          \mu_{ijk}^{\left(s\right)},
          \boldsymbol{\theta}_{j}^{\left(s\right)}
        \right)
      \right].
    \end{equation}
    For each model, we collect these contributions into an $S\times N_{\mathrm{obs}}$ matrix
    \begin{equation}
      \boldsymbol{\ell}
      =
      \left(\ell_{n}^{\left(s\right)}\right),
      \qquad
      s=1,\ldots,S,
      \qquad
      n=1,\ldots,N_{\mathrm{obs}},
    \end{equation}
    where $N_{\mathrm{obs}}$ is the total number of observations across all subjects and arms, and the single index $n$ enumerates the triples $\left(i,j,k\right)$.

    Given $\boldsymbol{\ell}$, PSIS-LOO provides an importance-sampling approximation to the LOO expected log predictive density, which we denote by $\widehat{\mathrm{elpd}}_{\mathrm{loo}}$, and the associated LOO information criterion $\mathrm{LOOIC} = -2\widehat{\mathrm{elpd}}_{\mathrm{loo}}$ for each model.

    \subsection{Posterior predictive residual checks}
    \label{app:ppc-residuals}

    Model adequacy was assessed using posterior predictive residual diagnostics based on the hierarchical data-generating mechanism implied by the fitted model. For each retained posterior draw $s=1,\ldots,S$ and each arm $j$, we first generated a new set of subject-specific parameters from the random effects and changepoint hierarchies, then simulated replicated latent outcomes $Y_{ijk}^{\ast,\mathrm{rep}\left(s\right)}$ from the skew-$t$ observation model. The same LLOQ $u_{ijk}$ was then applied,
    \begin{equation}
      Y_{ijk}^{\mathrm{rep}\left(s\right)}
      =
      \max\left\{
        Y_{ijk}^{\ast,\mathrm{rep}\left(s\right)},
        u_{ijk}
      \right\},
    \end{equation}
    so that the replicated data share the same observed-scale censoring mechanism as the original measurements.

    The collection of replicated datasets $\left\{Y_{ijk}^{\mathrm{rep}\left(s\right)}\right\}$ was used to construct simulation-based residuals via the \texttt{DHARMa} package \citep{HARTIG2021A,HARTIG2021B}. For each model, DHARMa returns scaled residuals that should be approximately $\mathrm{Uniform}\left(0,1\right)$ under correct posterior predictive calibration. We summarize these residuals using QQ plots that compare their empirical quantiles with the $\mathrm{Uniform}\left(0,1\right)$ reference distribution.

    As a complementary check, we constructed randomized probability integral transform (PIT) residuals from the hierarchical posterior predictive distributions on the observed scale, with randomization applied to account for the point mass induced by left censoring \citep{rosenblatt1952remarks, dunn1996randomized}. Index the $N_{\mathrm{obs}}$ observations by a single index $n$ enumerating $\left(i,j,k\right)$, and write $Y_{n}$ for the observed value with replicated values $Y_{n}^{\mathrm{rep}\left(s\right)}$, $s=1,\ldots,S$. For each $n$, we define the empirical lower and upper predictive distribution functions
    \begin{equation}
      \widehat{F}_{n}^{-}\left(y\right)
      =
      \frac{1}{S}
      \sum_{s=1}^{S}
      \mathbb{I}\left\{
        Y_{n}^{\mathrm{rep}\left(s\right)} < y
      \right\},
      \qquad
      \widehat{F}_{n}\left(y\right)
      =
      \frac{1}{S}
      \sum_{s=1}^{S}
      \mathbb{I}\left\{
        Y_{n}^{\mathrm{rep}\left(s\right)} \le y
      \right\},
    \end{equation}
    and construct the randomized PIT value
    \begin{equation}
      U_{n}
      =
      \widehat{F}_{n}^{-}\left(Y_{n}\right)
      +
      V_{n}
      \left\{
        \widehat{F}_{n}\left(Y_{n}\right)
        -
        \widehat{F}_{n}^{-}\left(Y_{n}\right)
      \right\},
      \qquad
      V_{n} \sim \mathrm{Uniform}\left(0,1\right),
      \label{eq:pit-randomized}
    \end{equation}
    independently across $n$. Because the censoring rule $Y_{n} = \max\left\{Y_{n}^{\ast},u_{n}\right\}$ is applied both to the observed data and to the replicated data, this PIT construction targets the posterior predictive distribution of the observed (censored) data. Under correct calibration, the collection $\left\{U_{n}\right\}$ is approximately $\mathrm{Uniform}\left(0,1\right)$, and the transformed residuals $Z_{n} = \Phi^{-1}\left(U_{n}\right)$ are approximately standard normal.

    \setlength{\bibhang}{1.5em}
    

    \newpage
    \clearpage

    \singlespacing

    \markboth{}{}

    \setcounter{page}{1}
    \setcounter{figure}{0}
    \setcounter{table}{0}
    \renewcommand\appendixname{Supplementary Material}
    \renewcommand{\figurename}{Supplementary Figure}
    \renewcommand{\tablename}{Supplementary Table}
    
    \begin{appendices}
    
        \maketitle
        
        \vskip 1.2truecm
        
        \begin{center}
            \noindent
            {\Large\bf Supplementary Material}
        \end{center}

        \newpage
        \clearpage

        \begin{longtable}{l S[table-format=4.3] S[table-format=3.3] S[table-format=2.3] l S[table-format=4.3] S[table-format=4.3] S[table-format=2.3]}
            \caption{Posterior summaries for model-level parameters and derived summaries under the abrupt and smooth changepoint models. Columns report the posterior median (Est.) and 95\% Bayesian credible interval (BCI). The quantities $\mu_{\tau,1}$ and $\sigma_{\tau,1}$ are the pre-truncation location and scale parameters for the subject-level changepoint distribution, and $\mu_{\kappa,1}$ and $\sigma_{\kappa,1}$ are the corresponding pre-truncation location and scale parameters for the subject-level transition sharpness distribution. The quantities $\bar{\tau}_{1}$, $\bar{\kappa}_{1}$, $\mathrm{SD}\left(\tilde{\tau}_{1}\right)$, and $\mathrm{SD}\left(\tilde{\kappa}_{1}\right)$ are the derived means and standard deviations of the fitted truncated normal distributions. The quantities $\mathrm{SD}\left(\alpha_{0i1}\right)$, $\mathrm{SD}\left(\beta_{1i1}\right)$, and $\mathrm{SD}\left(\beta_{2i1}\right)$ are the derived standard deviations of the random effects from $\boldsymbol{\Sigma}_{v,1}$, and $\rho_{\alpha_0,\beta_1}$, $\rho_{\alpha_0,\beta_2}$, and $\rho_{\beta_1,\beta_2}$ are the corresponding pairwise random effect correlations.}
            \label{tab:posterior-global}\\
            \toprule
            & \multicolumn{7}{c}{{\bf Changepoint Model}} \\
            \cline{2-8}
            & \multicolumn{3}{c}{{\bf Abrupt}} & & \multicolumn{3}{c}{{\bf Smooth}} \\
            \cline{2-4} \cline{6-8}
            \textbf{Parameter} &
            \textbf{Est.} &
            \multicolumn{2}{c}{{\bf 95\% BCI}} &
            &
            \textbf{Est.} &
            \multicolumn{2}{c}{{\bf 95\% BCI}} \\
            \midrule
            \endfirsthead
            
            \multicolumn{8}{l}{\textit{(Continued from previous page)}}\\
            \toprule
             & \multicolumn{7}{c}{{\bf Changepoint Model}} \\
            \cline{2-8}
            & \multicolumn{3}{c}{{\bf Abrupt}} & & \multicolumn{3}{c}{{\bf Smooth}} \\
            \cline{2-4} \cline{6-8}
            \textbf{Parameter} &
            \textbf{Est.} &
            \multicolumn{2}{c}{{\bf 95\% BCI}} &
            &
            \textbf{Est.} &
            \multicolumn{2}{c}{{\bf 95\% BCI}} \\
            \midrule
            \endhead
            
            \toprule
            \multicolumn{8}{r}{\textit{(Continued on next page)}}\\
            \endfoot
            
            \bottomrule
            \endlastfoot
            
              $\alpha_{0,1}$ & 5.092 & [4.909; & 5.274] &  & 5.099 & [4.921; & 5.275] \\ 
              $\beta_{1,1}$ & -3.750 & [-4.107; & -3.441] &  & -4.697 & [-5.542; & -4.102] \\ 
              $\beta_{2,1}$ & 3.484 & [3.158; & 3.841] &  & 4.563 & [3.928; & 5.423] \\ 
              $\beta_{1,1}+\beta_{2,1}$ & -0.267 & [-0.419; & -0.123] &  & -0.137 & [-0.310; & 0.026] \\ 
              $\mu_{\tau,1}$ & 0.530 & [0.467; & 0.593] &  & 0.436 & [0.341; & 0.517] \\ 
              $\sigma_{\tau,1}$ & 0.099 & [0.033; & 0.154] &  & 0.093 & [0.043; & 0.147] \\ 
              $\bar{\tau}_{1}$ & 0.530 & [0.467; & 0.593] &  & 0.436 & [0.342; & 0.517] \\ 
              $\mathrm{SD}\left(\tilde{\tau}_{1}\right)$ & 0.099 & [0.033; & 0.154] &  & 0.093 & [0.043; & 0.146] \\ 
              $\mu_{\kappa,1}$ &  &  &  &  & 3.608 & [-11.340; & 6.909] \\ 
              $\sigma_{\kappa,1}$ &  &  &  &  & 3.377 & [1.263; & 13.450] \\ 
              $\bar{\kappa}_{1}$ &  &  &  &  & 4.789 & [2.898; & 9.484] \\ 
              $\mathrm{SD}\left(\tilde{\kappa}_{1}\right)$ &  &  &  &  & 2.775 & [1.225; & 6.880] \\ 
              $\sigma_{1}$ & 0.624 & [0.411; & 1.461] &  & 0.463 & [0.331; & 1.043] \\ 
              $\Lambda_{1}$ & 0.253 & [0.003; & 0.482] &  & 0.209 & [-0.034; & 0.449] \\ 
              $\nu_{1}$ & 2.323 & [2.047; & 3.148] &  & 2.542 & [2.077; & 3.836] \\ 
              $\mathrm{SD}\left(\alpha_{0i1}\right)$ & 0.544 & [0.431; & 0.700] &  & 0.540 & [0.428; & 0.693] \\ 
              $\mathrm{SD}\left(\beta_{1i1}\right)$ & 0.290 & [0.077; & 0.735] &  & 0.257 & [0.074; & 0.642] \\ 
              $\mathrm{SD}\left(\beta_{2i1}\right)$ & 0.393 & [0.083; & 0.936] &  & 0.384 & [0.095; & 0.887] \\ 
              $\rho_{\alpha_0,\beta_1}$ & 0.101 & [-0.704; & 0.911] &  & -0.299 & [-0.942; & 0.832] \\ 
              $\rho_{\alpha_0,\beta_2}$ & -0.214 & [-0.901; & 0.532] &  & 0.185 & [-0.697; & 0.823] \\ 
              $\rho_{\beta_1,\beta_2}$ & -0.528 & [-0.970; & 0.909] &  & -0.129 & [-0.956; & 0.949] \\ 
        \end{longtable}

        \begin{landscape}
        
        \begin{figure}[!htbp]
            \centering
            \includegraphics[
                page=1,
                width=0.95\linewidth,
                height=0.83\textheight,
                keepaspectratio
            ]{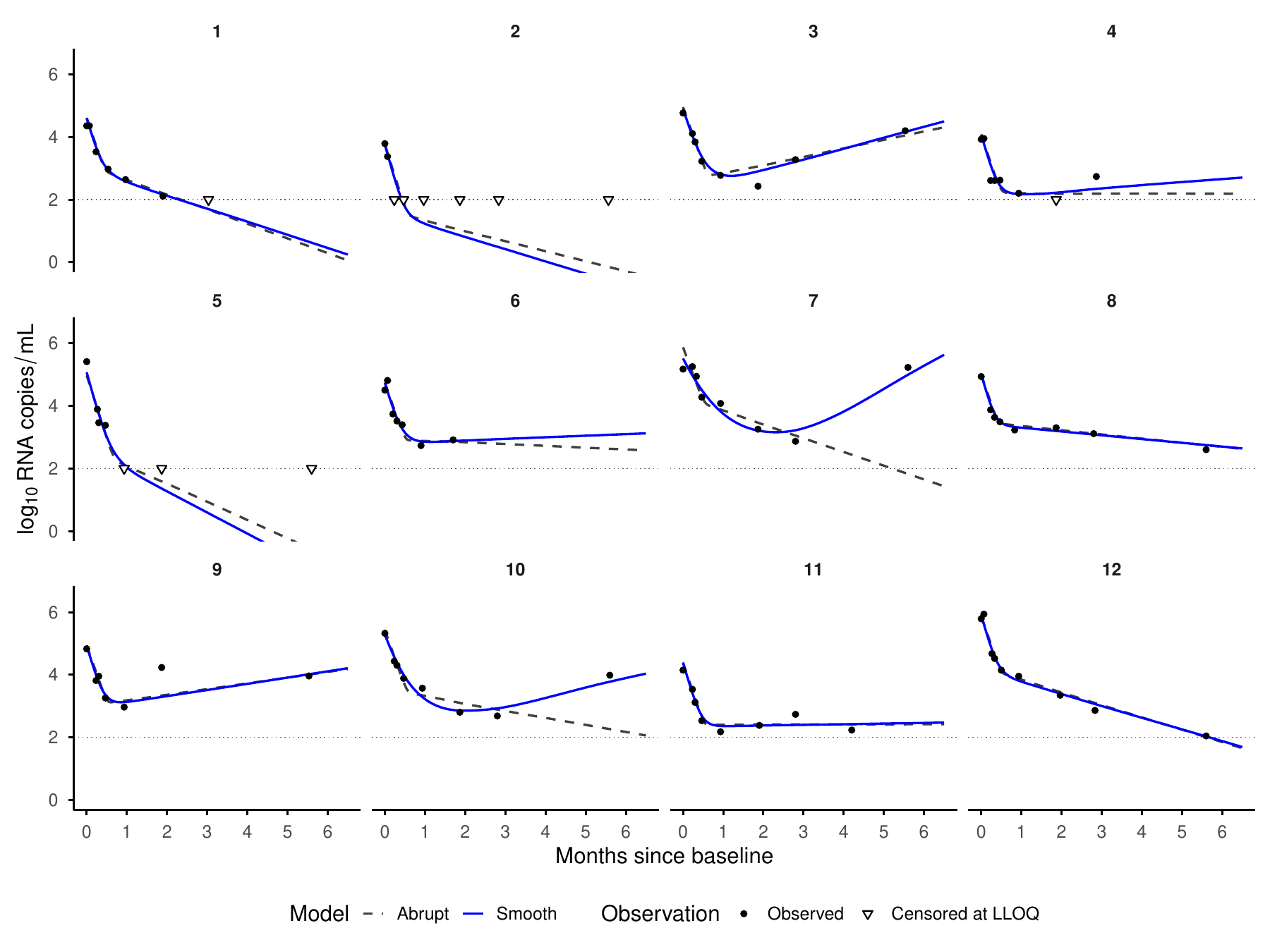}
            \caption{Subject-specific fitted trajectories under the abrupt and smooth changepoint models for the ACTG~315 data. Points show observed $\log_{10}$ HIV-1 RNA values, and open triangles show measurements censored at the LLOQ, plotted at $\log_{10}\left(100\right)=2$. Lines show pointwise posterior medians of the fitted latent mean trajectories under each model, with panels indexed by participant. Part~1 of 4.}
            \label{fig:patient-fits}
        \end{figure}
        
        \begin{figure}[!htbp]
            \ContinuedFloat
            \centering
            \includegraphics[
                page=2,
                width=0.95\linewidth,
                height=0.83\textheight,
                keepaspectratio
            ]{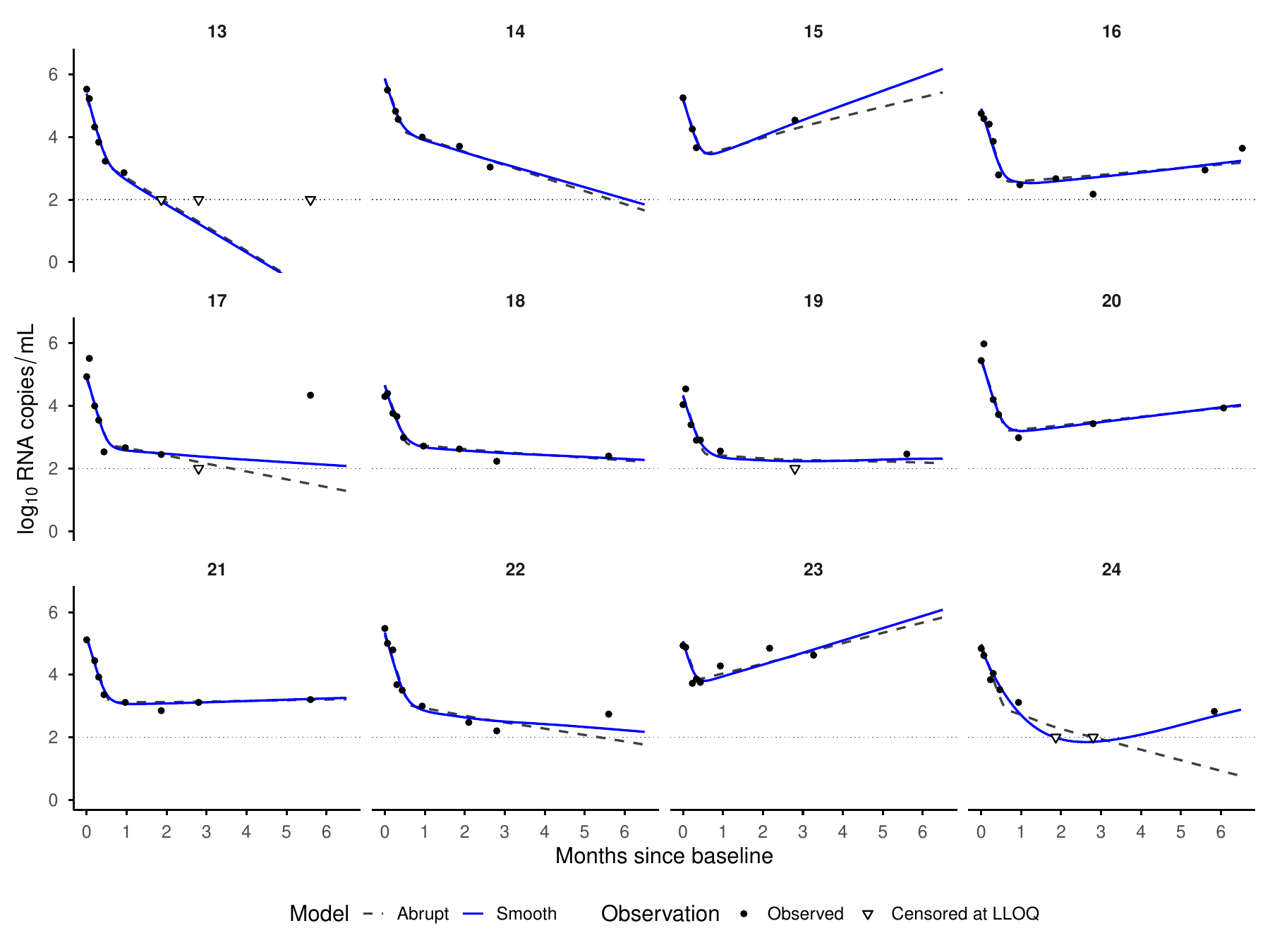}
            \caption[]{Subject-specific fitted trajectories under the abrupt and smooth changepoint models, continued. Part~2 of 4.}
        \end{figure}
        
        \begin{figure}[!htbp]
            \ContinuedFloat
            \centering
            \includegraphics[
                page=3,
                width=0.95\linewidth,
                height=0.83\textheight,
                keepaspectratio
            ]{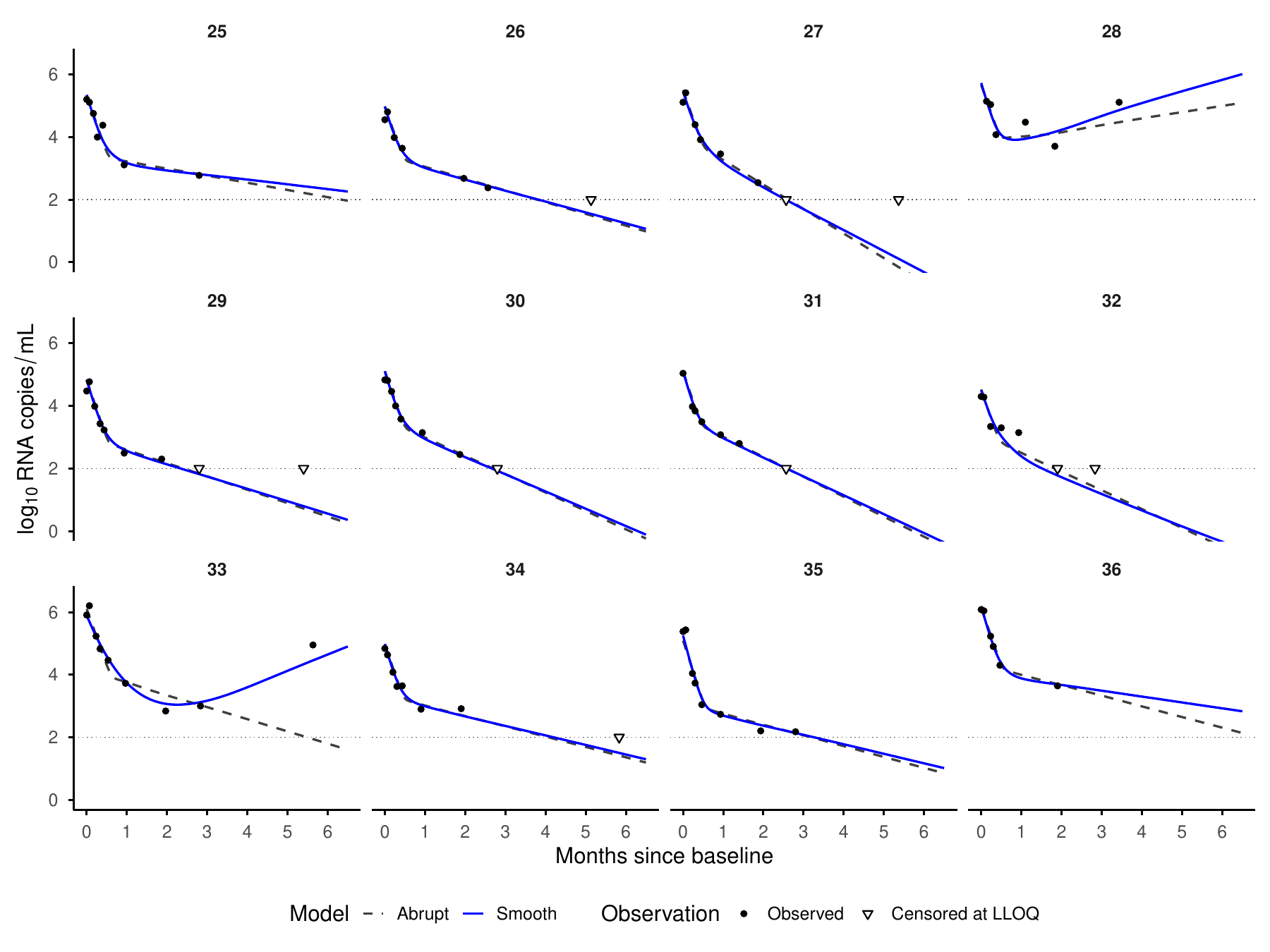}
            \caption[]{Subject-specific fitted trajectories under the abrupt and smooth changepoint models, continued. Part~3 of 4.}
        \end{figure}
        
        \begin{figure}[!htbp]
            \ContinuedFloat
            \centering
            \includegraphics[
                page=4,
                width=0.95\linewidth,
                height=0.83\textheight,
                keepaspectratio
            ]{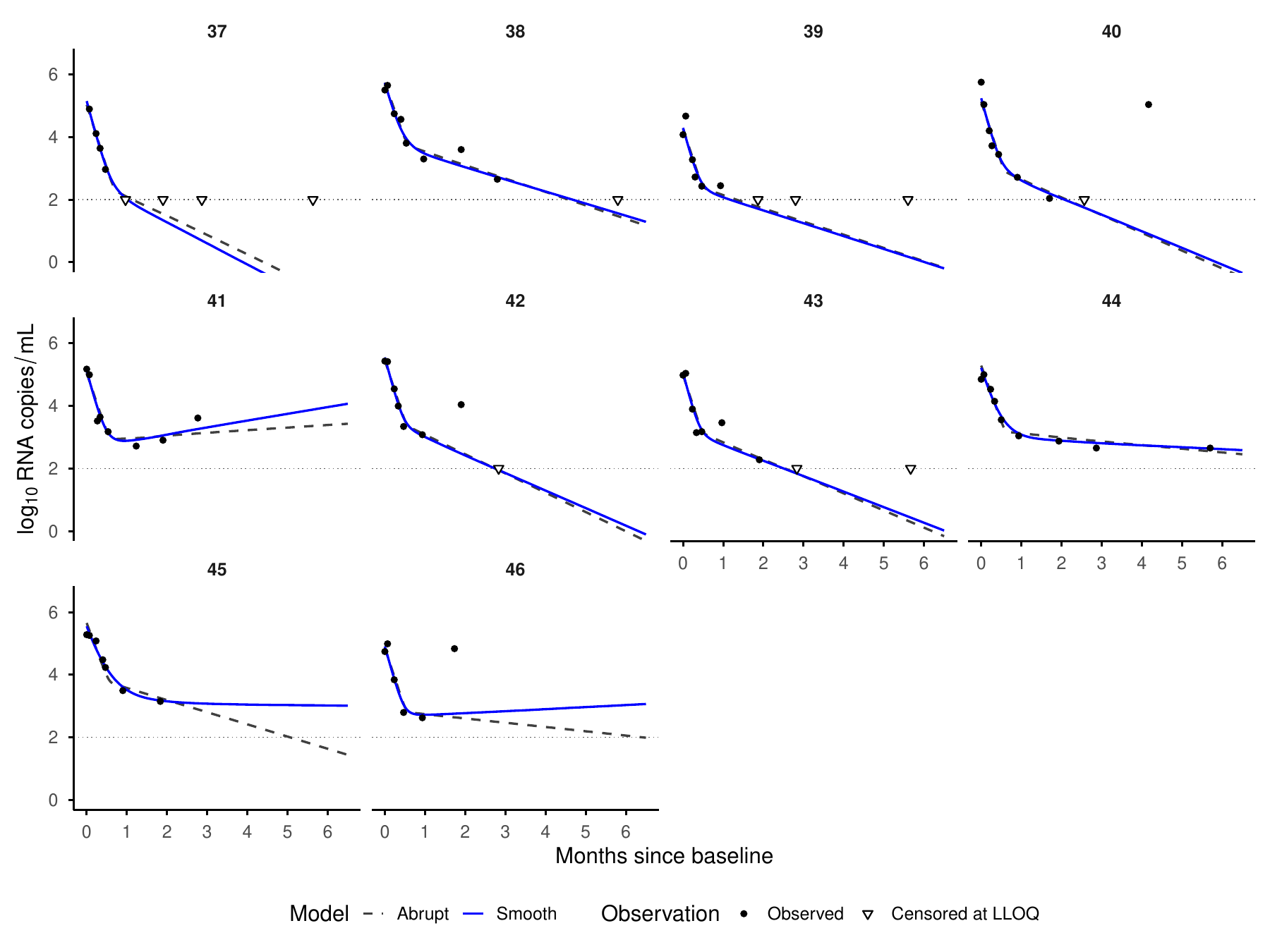}
            \caption[]{Subject-specific fitted trajectories under the abrupt and smooth changepoint models, continued. Part~4 of 4.}
        \end{figure}
        
        \end{landscape}

        \begin{figure}[!htbp]
            \centering
            
            \begin{subfigure}{0.69\textwidth}
                \centering
                \includegraphics[width=\linewidth]{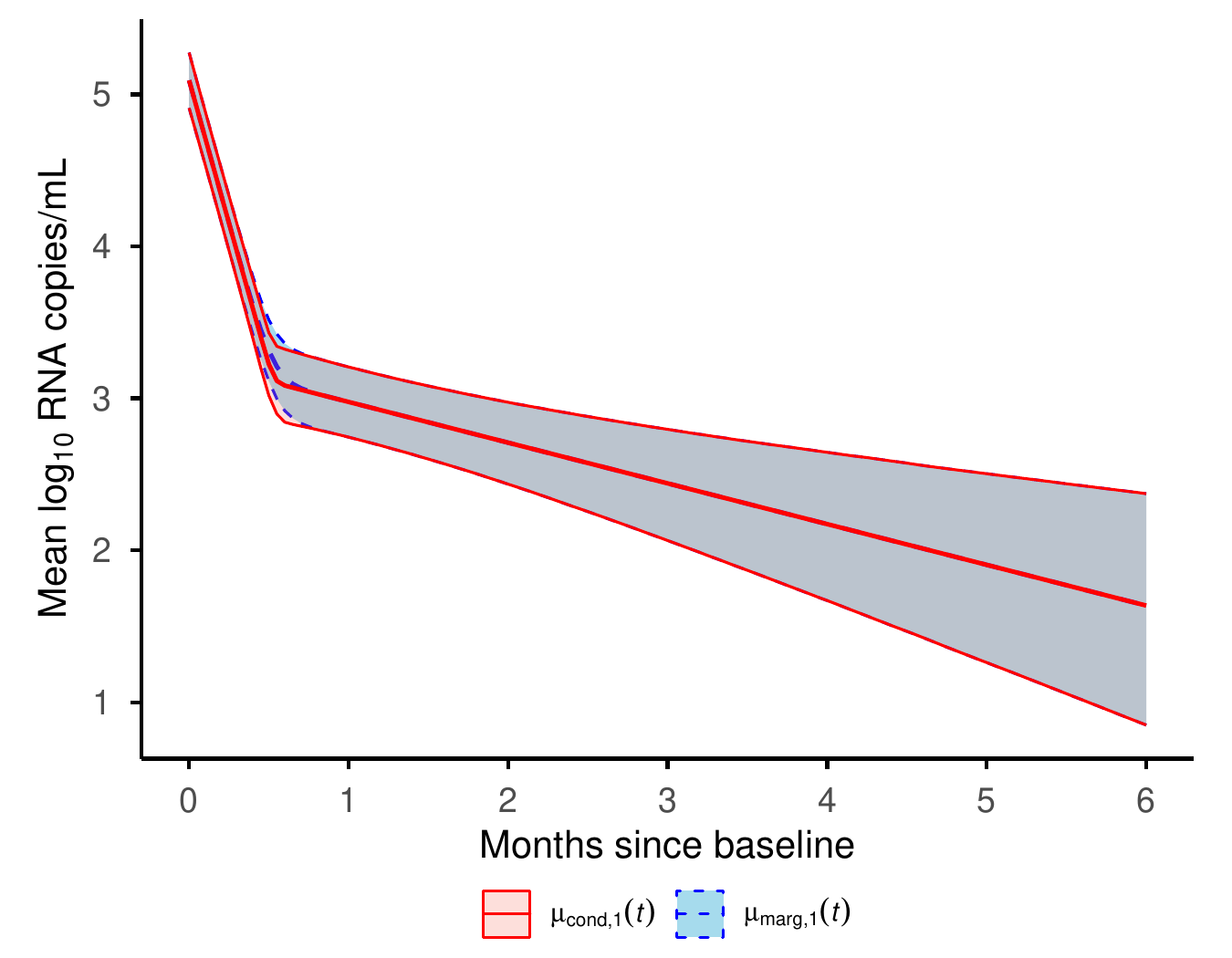}
                \caption{Mean trajectory}
                \label{fig:bilinear-mean}
            \end{subfigure}
                
            \vspace{1em}
                
            \begin{subfigure}{0.69\textwidth}
                \centering
                \includegraphics[width=\linewidth]{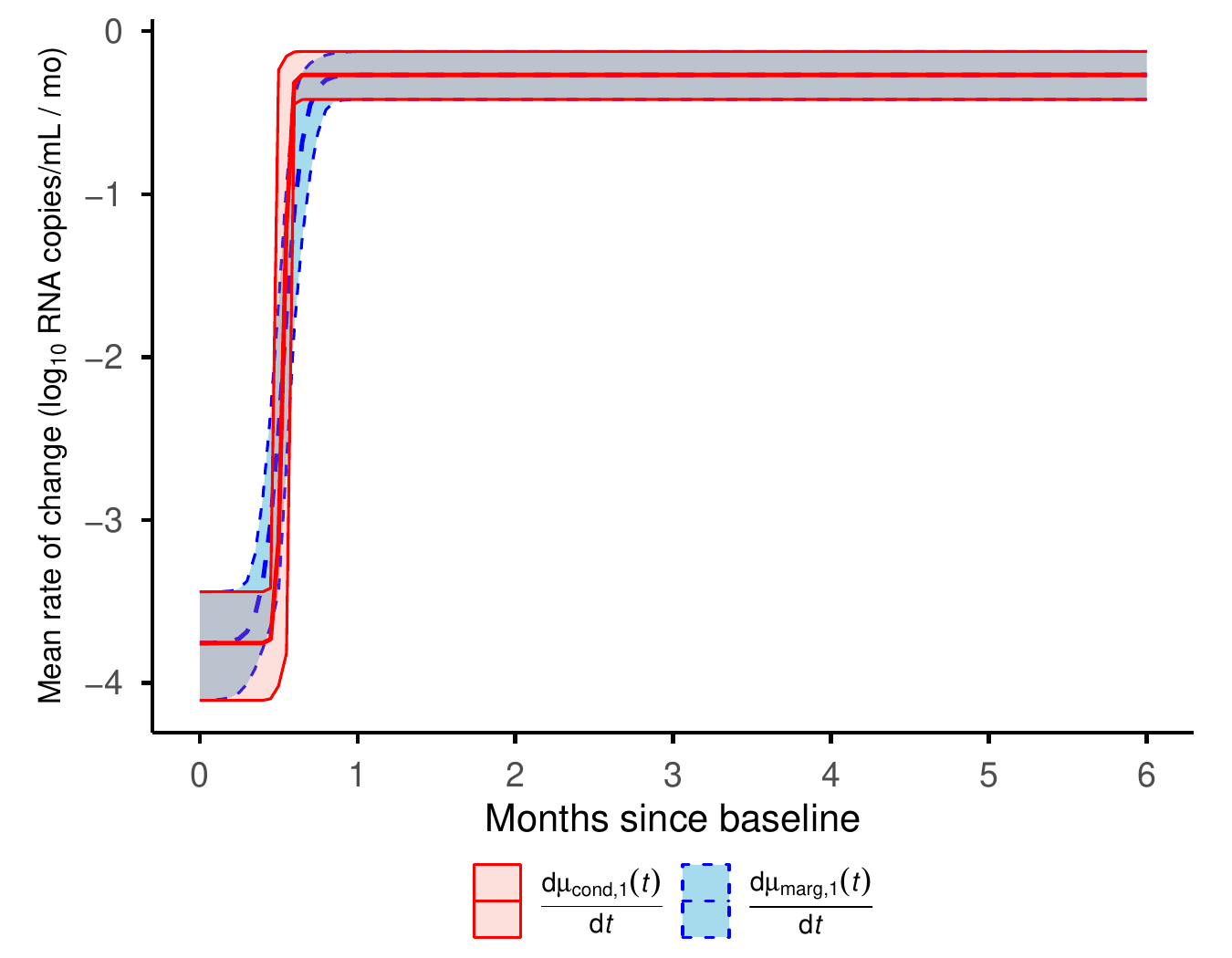}
                \caption{Instantaneous slope}
                \label{fig:bilinear-slope}
            \end{subfigure}
            
            \caption{Abrupt changepoint model. Panel (a) shows the fitted mean trajectory for the underlying $\log_{10}$ HIV-1 RNA process over time; panel (b) shows the instantaneous rate of change in $\log_{10}$ HIV-1 RNA (copies/mL per month). In each panel, the dashed curve and band are the marginal summaries, and the solid curve and band are the corresponding conditional (plug-in) summaries.}
        \label{fig:bilinear-main}
        \end{figure}
        
        \begin{figure}[!htbp]
            \centering
            \begin{subfigure}{0.49\textwidth}
                \centering
                \includegraphics[width=\linewidth]{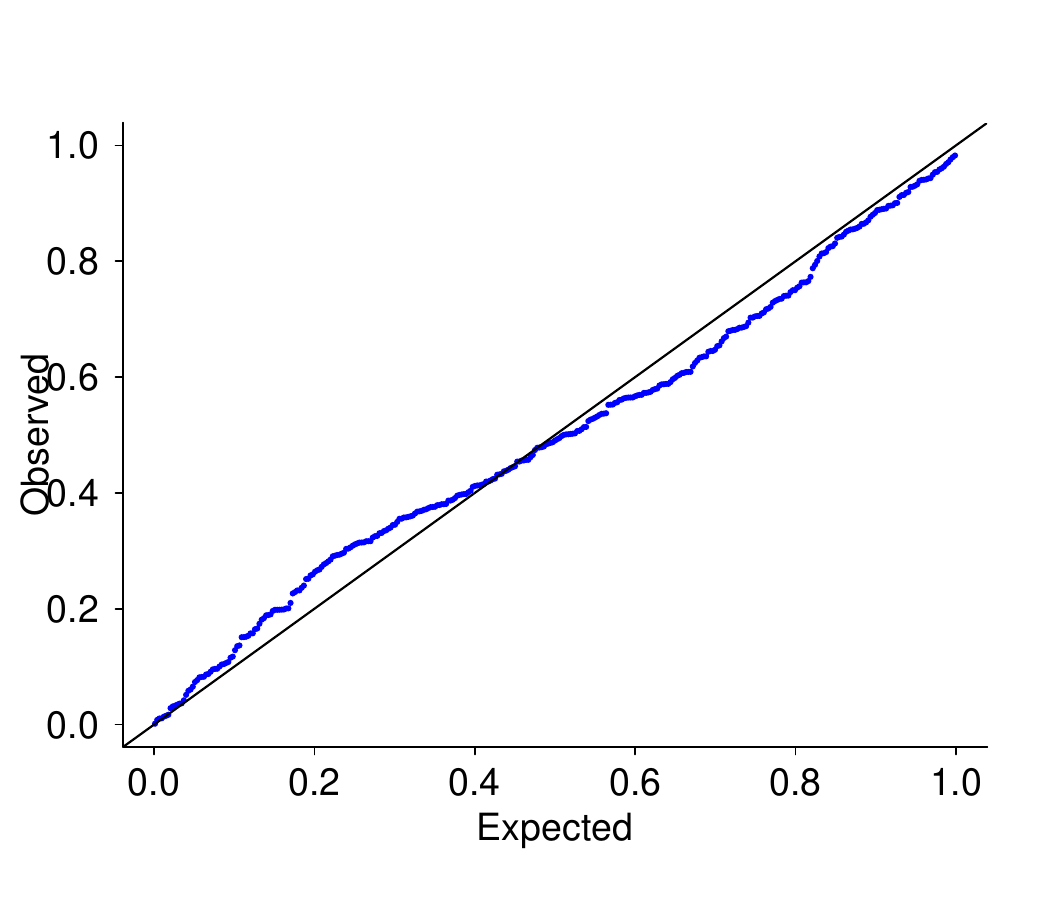}
                \caption{Abrupt changepoint model}
            \end{subfigure}
            \begin{subfigure}{0.49\textwidth}
                \centering
                \includegraphics[width=\linewidth]{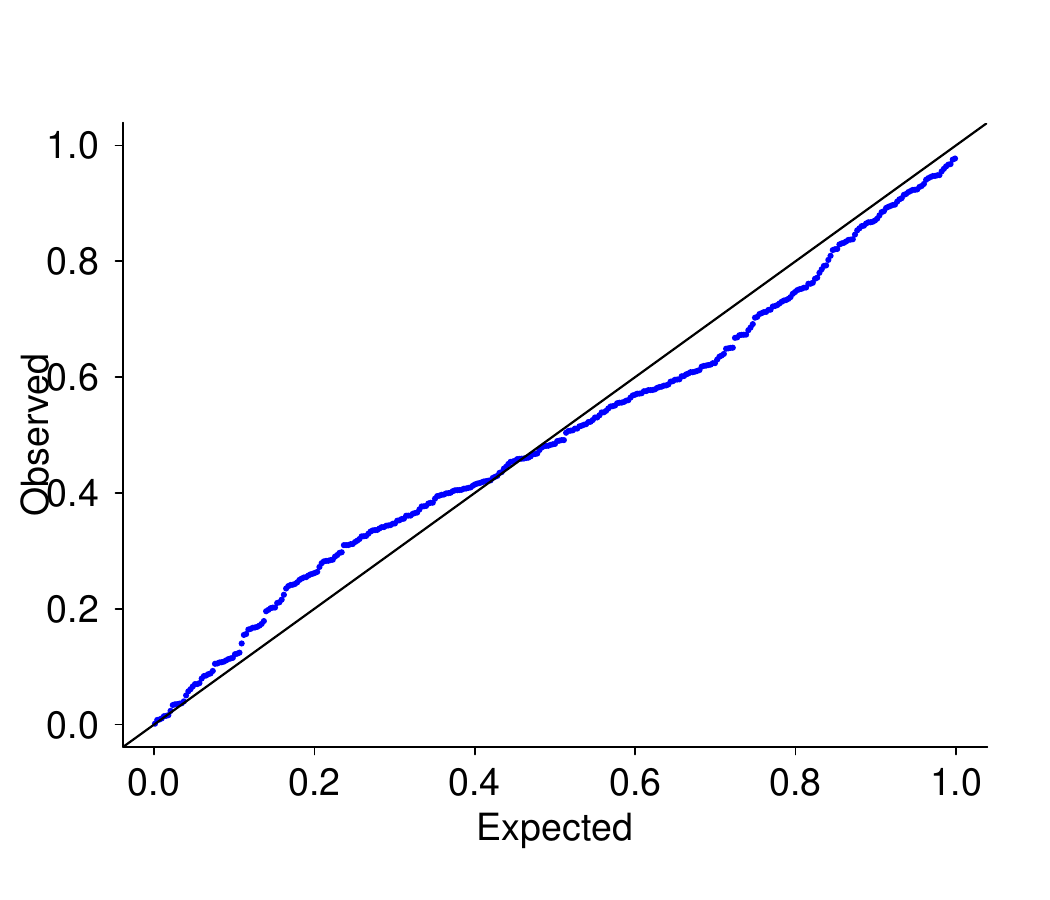}
                \caption{Smooth changepoint model}
            \end{subfigure}
            \caption{DHARMa simulation-based QQ plots for posterior predictive residuals. Expected quantiles correspond to a $\mathrm{Uniform}\left(0,1\right)$ distribution, and observed quantiles correspond to the empirical quantiles of the DHARMa scaled residuals.}
            \label{fig:ppc-dharma-qq}
        \end{figure}
    
        \begin{figure}[!htbp]
            \centering
            \begin{subfigure}{0.49\textwidth}
                \centering
                \includegraphics[width=\linewidth]{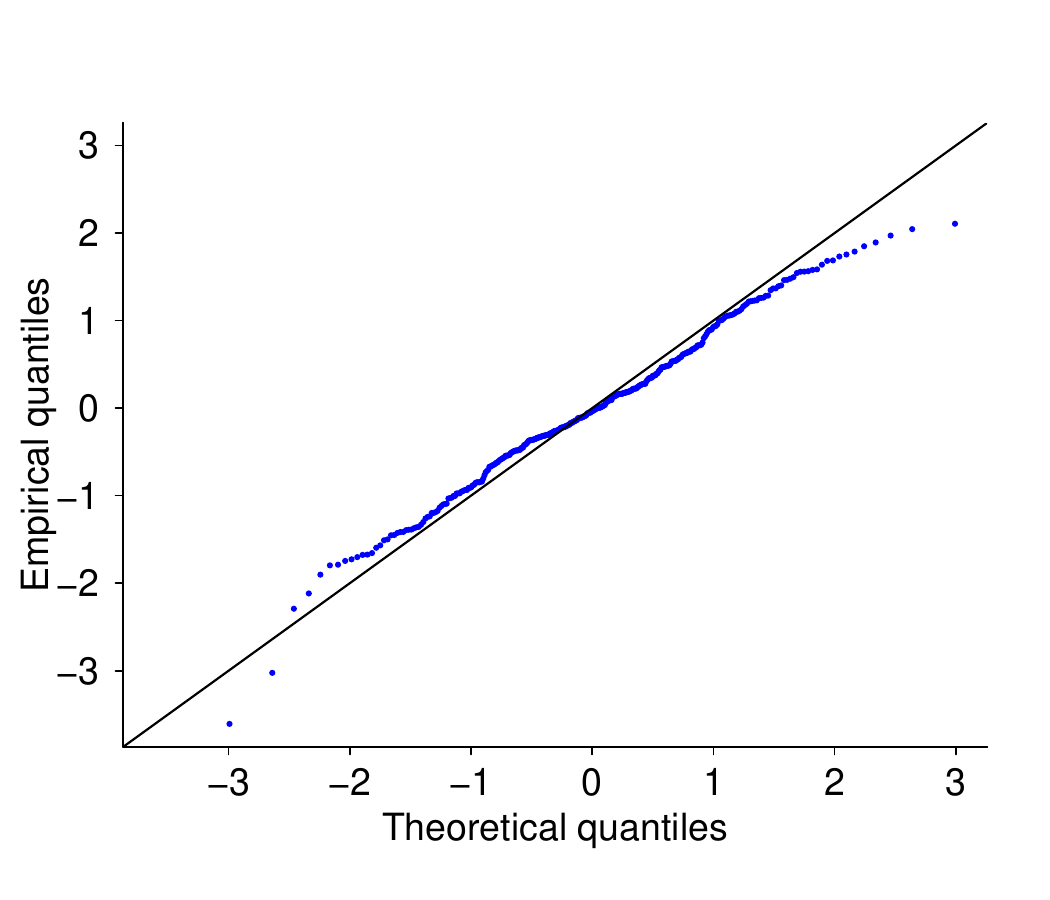}
                \caption{Abrupt changepoint model}
            \end{subfigure}
            \begin{subfigure}{0.49\textwidth}
                \centering
                \includegraphics[width=\linewidth]{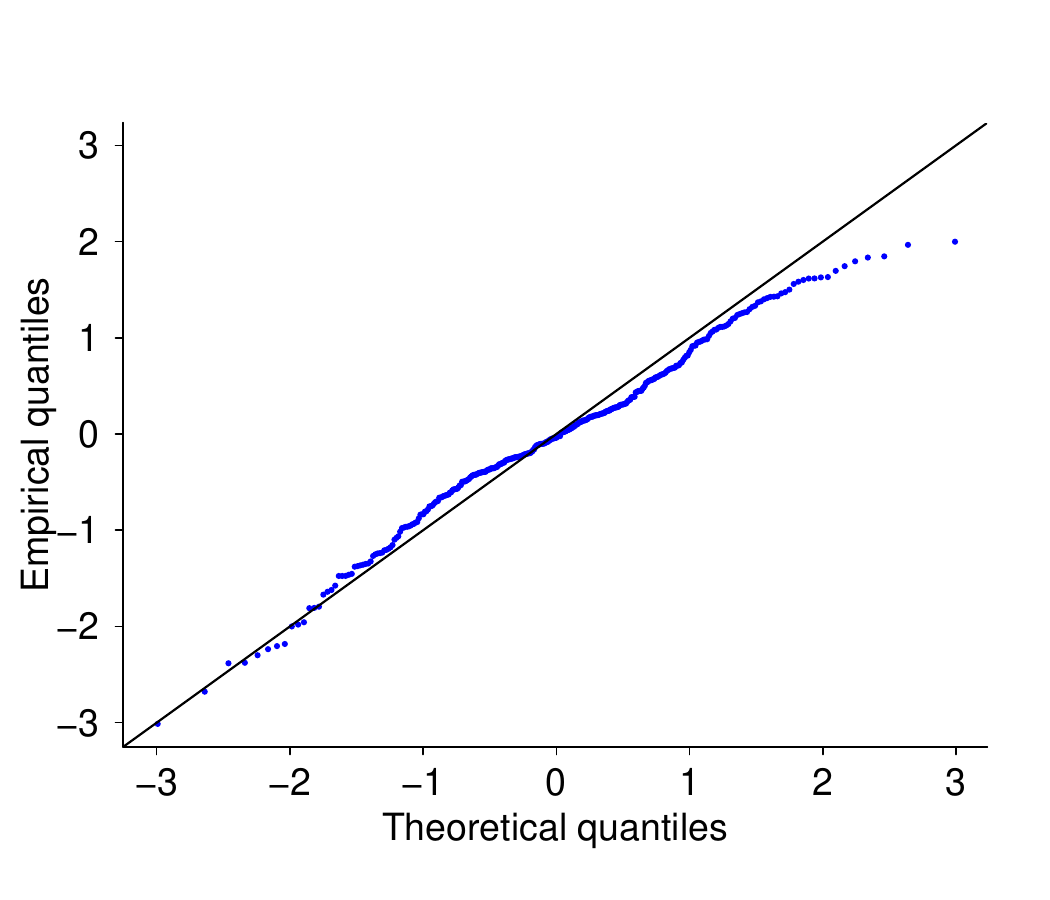}
                \caption{Smooth changepoint model}
            \end{subfigure}
            \caption{Normal QQ plots of the randomized PIT residuals $Z_{n} = \Phi^{-1}\left(U_{n}\right)$. Under correct calibration, the points should lie close to the 45-degree reference line.}
            \label{fig:ppc-pit-qq}
        \end{figure}

    \end{appendices}

\end{document}